\documentclass[usenatbib]{mnras}
\usepackage[utf8]{inputenc}
\usepackage{natbib,hyperref,pdflscape}
\usepackage{graphicx,mwe,subcaption}
\usepackage[export]{adjustbox}

\title[MYSO multiplicity in the $K-$band]{The RMS survey: a census of massive YSO multiplicity in the $K-$band}
\author[R. G. Shenton et al.]{Robert G. Shenton$^{1}$\thanks{E-mail: py15rgs@leeds.ac.uk}, Rebecca J. Houghton$^{2}$, René D. Oudmaijer$^{1}$, Simon P. Goodwin$^{2}$,\and Stuart L. Lumsden$^{1}$, Evgenia Koumpia$^{3}$ and Maria Koutoulaki$^{1}$\\\\$^{1}$School of Physics and Astronomy, University of Leeds, Leeds, LS2 9JT, UK\\$^{2}$Department of Physics and Astronomy, University of Sheffield, Sheffield, S3 7RH, UK\\$^{3}$ESO, Alonso de Córdova 3107 Vitacura, Casilla, 19001, Santiago, Chile}
\date{Accepted XXX. Received YYY; in original form ZZZ}
\pubyear{2023}

\begin{document}
\bibliographystyle{mnras}

\maketitle

\begin{abstract}
Close to 100 per cent of massive stars are thought to be in binary systems. The multiplicity of massive stars seems to be intrinsically linked to their formation and evolution, and Massive Young Stellar Objects are key in observing this early stage of star formation. We have surveyed three samples totalling hundreds of MYSOs ($>8M_\odot$) across the Galaxy from the RMS catalogue, using UKIDSS and VVV point source data, and UKIRT $K-$band imaging to probe separations between 0.8-9 arcsec (approx 1000-100,000 au). We have used statistical methods to determine the binary statistics of the samples, and we find binary fractions of $64\pm 4$ per cent for the UKIDSS sample, $53\pm 4$ per cent for the VVV sample, and $49\pm 8$ per cent for the RMS imaging sample. Also we use the $J-$ and $K-$band magnitudes as a proxy for the companion mass, and a significant fraction of the detected systems have estimated mass ratios greater than 0.5, suggesting a deviation from the capture formation scenario which would be aligned with random IMF sampling. Finally, we find that YSOs located in the outer  Galaxy have a higher binary fraction than those in the inner Galaxy. This is likely due to a lower stellar background density than observed towards the inner Galaxy, resulting in higher probabilities for visual binaries to be physical companions.  It does indicate a binary fraction in the probed separation range of close to 100 per cent without the need to consider selection biases.

\end{abstract}


\section{Introduction}
Star formation has been an intense point of research in recent years, however the formation of massive ($>8 M_\odot$) stars is still not fully understood. A crucial part of the debate revolves around the question whether the formation scenario for massive stars is simply a variation of the intermediate and low-mass star formation theories, or whether they have a completely different origin. Stellar multiplicity has its own implications on the process of star formation, and multiplicity properties are established early on in the lives of stellar systems, particularly in the pre-main-sequence (PMS) stage \citep{MAT94, DUC13}. A large proportion of stars are thought to form in multiple systems \citep{DUC13}, and it is also known that up to 100 per cent of OB-type stars are in multiple systems \citep{CHI12}. Multiplicity also significantly affects the ongoing evolution of massive stars \citep{SAN12} and may trigger further star formation through outflows, which makes them a significant factor in the evolution of galaxies and the interstellar medium \citep{KEN05}.

\cite{DUC13} and \cite{OFF23} review the theories on how massive stars may form in binary and multiple systems. Two of the most favoured formation scenarios are disc fragmentation and capture. Disc fragmentation - where the accretion disc around a prestellar core experiences gravitational instability and fragments into clumps - is more common for massive stars, as gravitational instability is more likely to occur in massive systems than low-mass systems \citep{KRA08,KRU09}. In binary capture, two isolated stars form and then interact to become a gravitationally bound pair. Both of these formation scenarios predict close ($<$ 100 au) binaries through simulations \citep{MEY18}, whereas larger separations are predicted by simulations due to fragmentation processes during the collapse phase instead \citep{MYE13}. \cite{KIN12} and \cite{MAR12} suggest that multiplicity tends to be higher in denser clusters.

Studies have shown how different factors in massive star formation can affect binarity. The multiplicity of a system has been shown to scale with the mass of the primary object \citep{OFF23}. Magnetic fields and radiative feedback may prevent the fragmentation process from becoming too violent, which would cause stellar ejections and reduce the overall multiplicity fraction \citep{BAT12}.  Later on in the PMS phase, accretion disks form around massive stars and eventually fragment to form companions \citep{ROS19}. Primordial massive wide binaries (MWBs) with separations $>10^2$ au are more likely to survive in low-density regions with few surrounding stars; in high density regions they have a high risk of destruction \citep{GRI18}. Ultra-wide binaries at even larger separations ($10^4 - 10^5$ au) are known to exist due to cluster evolution \citep{MOE10}. Meanwhile, massive close binaries may be a result of inward migration from wider separations, occurring through interaction with a disk remnant or another stellar object or the result of magnetic braking \citep{LUN18,RAM21, Harada2021}. However, Atacama Large Millimetre/submillimetre Array (ALMA) observations of high-mass star-forming regions have indicated that the fragmentation process occurs down at the smallest observable scales \citep{BEU19, MEY19}. 

53 per cent of massive main-sequence (MS) O- type stars have been reported to be in binary systems at separations less than 200 au, with the multiplicity fraction increasing to 90 per cent for larger separations \citep{SAN14, BOR22}. As a result of dynamical processes such as capture or magnetic braking \citep{LUN18} occurring during the evolution of a star, the multiplicity statistics of MS stars may not be an accurate indicator of the primordial properties of a multiple system \citep{KRA11}. In order to verify the theories suggested for pre-MS binary formation, observational studies of MYSOs (Massive Young Stellar Objects) play an important role.

MYSOs represent a key point early on in a star's lifetime where the process of accretion can be observed and investigated. This phase lasts around $10^5$ years, and heavy dust extinction is common during this phase which renders the majority of MYSOs effectively invisible at $<1\mu m$ \citep{DAV11}. They are bright in the mid-infrared which makes this wavelength range ideal for observing them. Small-scale gap-like substructures in MYSO disks have been connected to the high binary fractions of MYSOs and may be due to the presence of one or more companions \citep{FRO21}.

Little work has been done on MYSO multiplicity; however there has still been a significant number of reported binaries, of which a large fraction were anecdotal or serendipitous discoveries (e.g. \citealt{KRA17, KOU19, ZHA19, Cyganowski2022}). The closest spatially resolved MYSO binary systems were reported by \cite{KOU19}. These authors investigated two MYSOs using H-band VLTI/PIONIER observations and found companions at separations of 30 au for PDS 27, and 42-54 au for PDS 37 respectively. \cite{KOU21} presented the first interferometric K-band survey of MYSOs using VLTI observations of six objects, and found a low binary fraction of $17\pm15$ per cent at separations between 2-300 au. 

The first dedicated survey into the multiplicity of MYSOs comes from \cite{POM19} who analysed a sample of 32 objects from the RMS survey catalogue \citep{LUM13}. Using adaptive optics $K$-band observations, 18 previously undiscovered companions were discovered within 600-10,000 au of the primaries. The multiplicity fraction was found to be $31\pm8$ per cent and the companion fraction was reported to be $53\pm9$ per cent, although it was asserted that the true multiplicity fraction could be up to 100 per cent. Mass ratios for the sample were generally found to be greater than $0.5$, suggesting binary capture was not responsible for forming these systems. These results are consistent with multiplicity studies on the intermediate mass pre-main sequence Herbig AeBe stars \citep{BAI12, WHE10}. However, caveats of the survey include the small sample size, and the shallow limiting magnitude (between $K=12$ and $K=15$). This paper aims to further the work done by Pomohaci's pilot survey of 32 objects, using a much larger sample of hundreds of MYSOs.

This paper is structured as follows. Section 2 outlines the nature of the observations used in the sample of MYSOs. Section 3 explains the results of the multiplicity analysis, including the details of completeness and accounting for chance projections. In Section 4 we discuss the multiplicity statistics achieved from this sample and compare them to other previous studies, and we also explore mass ratios of the potential companions detected. Section 5 summarises our findings.

\section{Observational Data}
\subsection{Sample selection}

All of our targets are drawn from the Red MSX Source (RMS) survey \citep{LUM13}. This survey was constructed with the aim of creating a complete and unbiased database of the Galactic population of Young Stellar Objects (YSOs), by using multiwavelength data to discern YSOs from other similar objects, including HII regions and evolved stars. The full catalogue can be found at \url{http://rms.leeds.ac.uk}. The survey is complete for massive protostellar objects brighter than $2 \times 10^4 L_{\odot}$ out to 18 kpc, and is restricted to $10^{\circ}<l<350^{\circ}$ to avoid source confusion towards the Galactic centre. The YSOs in our sample have distances ranging between 1.4-11.2 kpc; for our chosen detection range of 0.5-9 arcsec, this places any detected companions between 700-100,000 au away from the primary. The YSOs have masses ranging from 1.9-8 M$_\odot$, while the MYSOs have masses ranging between 8-49.5 M$_\odot$.

\subsection{Galactic Plane Surveys}
Point source catalogue data from the UKIRT Infrared Deep Sky Survey Galactic Plane Survey (UKIDSS GPS, \citealt{UKIDSS}) was used to find targets in the Northern sky. The $K-$band was used so that YSOs are visible despite high extinction. The UKIRT Wide Field Camera (WFCAM) used for UKIDSS has a pixel size of 0.4", and the limiting magnitude of the data is $K$=19. The GPS survey has a spatial resolution of 0.8-1". In the UKIDSS DR11 catalogue, 395 YSOs were found, with 221 classed as MYSOs.

Alongside UKIDSS, point source catalogue data from the Vista Variables in the Via Lactea (VVV, \citealt{VVV}) survey was used. VVV focuses on the Southern part of the Galactic plane, and DR5 contains data on 279 YSOs, with 181 of them classed as MYSOs. The VVV DR5 catalogue does not cover the entirety of the Southern sky, and so there is a region of the galactic plane left uncovered by either of these surveys.
Additionally there is an overlap of two objects between UKIDSS and VVV for our YSO samples. The VISTA IR Camera (VIRCAM) used in VVV has a pixel size of 0.34" and an average limiting magnitude of $K_s$=18.5, with a spatial resolution of $\sim$0.9".

863 objects labelled as `YSO' or `HII/YSO' are present in the RMS catalogue. 681 of these were found in the UKIDSS/VVV surveys; the remainder were not found in either survey. The full table of YSOs can be found in \autoref{appA}. The main benefit of using these surveys is their coverage of the RMS catalogue, as well as their deep limiting magnitudes, and the availability of multi-colour data (specifically $J$- and $H$-bands) which is useful in determining interstellar extinction. These data allows deeper probing than the NaCo images used in \citet{POM19} which had an average limiting magnitude of $K$=14. The main trade-off of our study compared to NaCo is the relatively lower spatial resolution of these surveys.  In addition, the 2MASS survey was also used for photometry brighter than the saturation limit of UKIDSS/VVV ($K\sim 12$).
2MASS uses a pixel size of 1" and has a spatial resolution of $\sim$2", meaning it has only a quarter of the resolution of UKIDSS/VVV.

\subsection{UKIRT/RMS K-band imaging}

K-band imaging data was obtained for a sample of 88 RMS objects (referred to from here onwards as the `RMS images'), taken by the United Kingdom Infra-Red Telescope (UKIRT) in Hawaii between 2001 and 2006. 38 images were taken using the UKIRT 1-5 micron Imager Spectrometer (UIST) instrument and 50 were taken with the UKIRT Fast-Track Imager (UFTI) as a follow-up.

These 88 YSOs were randomly sampled from the RMS catalogue. The RMS images were acquisition images originally used for obtaining spectra \citep{CLA06,COOTHESIS}, and these images were calibrated using flat field frames and sky subtraction, and also had their astrometry corrected. The field of view of each of the images is $2.3$ arcminutes. The images have an average limiting magnitude of $K$=17.5, and a seeing of $\sim$0.7" on average. The UIST and UFTI instruments of UKIRT have pixel sizes of 0.12" and 0.09" respectively. The main benefit of these images is the improved resolution compared to UKIDSS/VVV. UKIDSS/VVV data were used as a reference to calibrate the $K$-band flux in the RMS images. There is no overlap with the VVV catalogue but 75 of the YSOs in the RMS image sample are also in our UKIDSS sample.

\begin{figure}
        \centering
        \begin{subfigure}[H]{0.9\columnwidth}
            \centering
            \includegraphics[width=0.98\columnwidth]{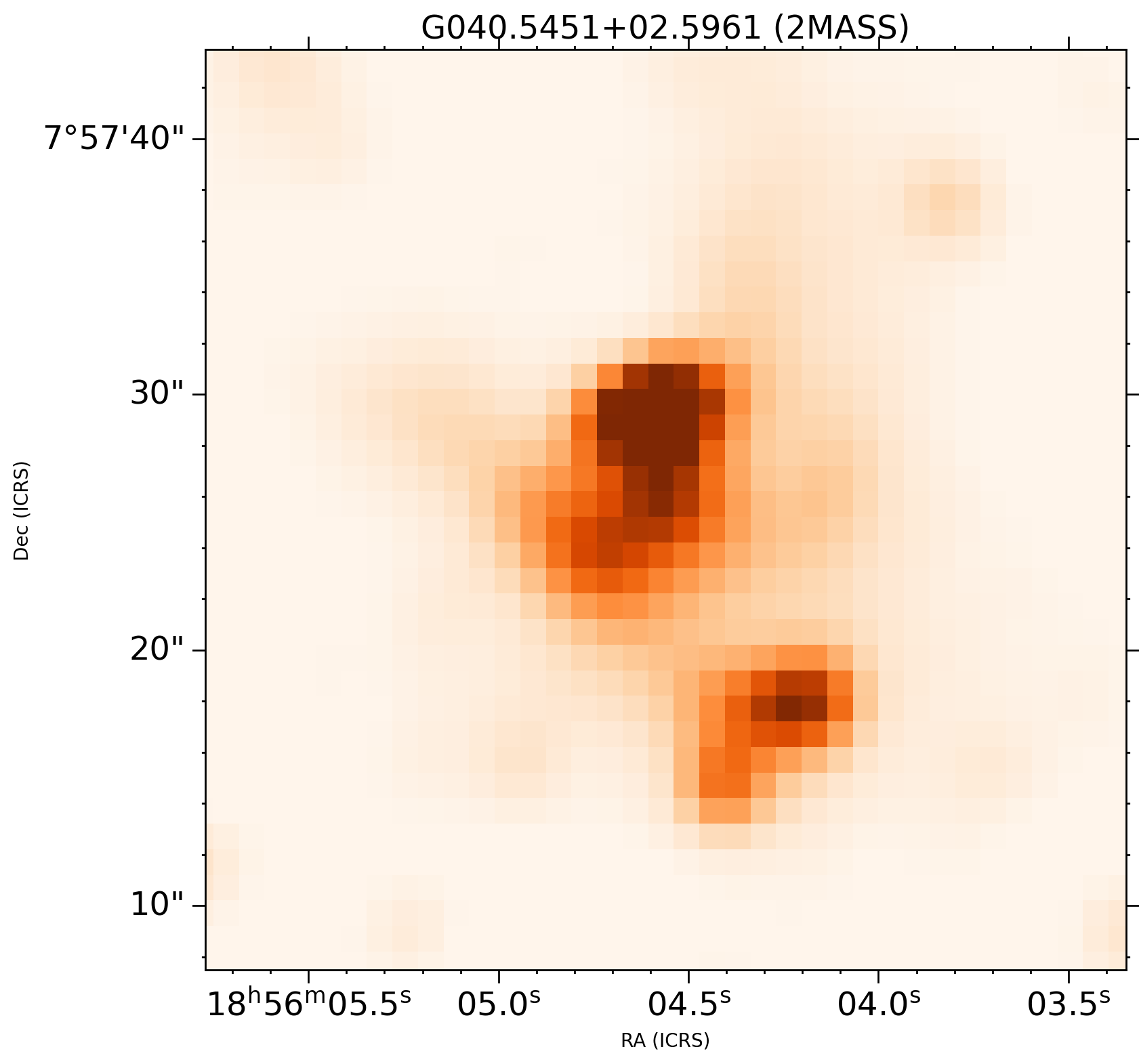}
            \label{fig:G040_2MASS}
        \end{subfigure} \\
        \begin{subfigure}[H]{0.9\columnwidth}  
            \centering
            \includegraphics[width=0.98\columnwidth]{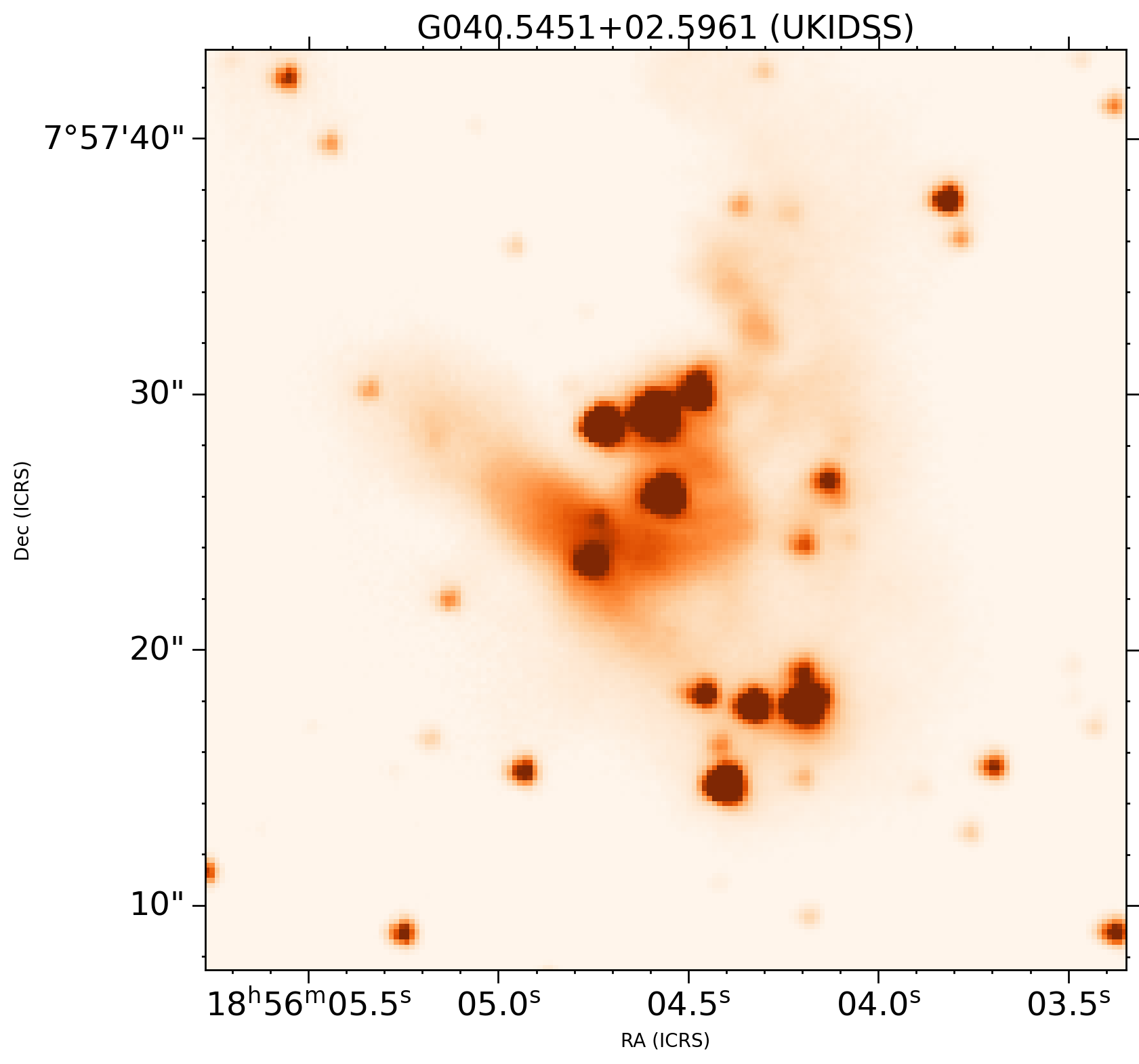}
            \label{fig:G040_UKIDSS}
        \end{subfigure} \\
        \begin{subfigure}[H]{0.9\columnwidth}   
            \centering
            \includegraphics[width=0.98\columnwidth]{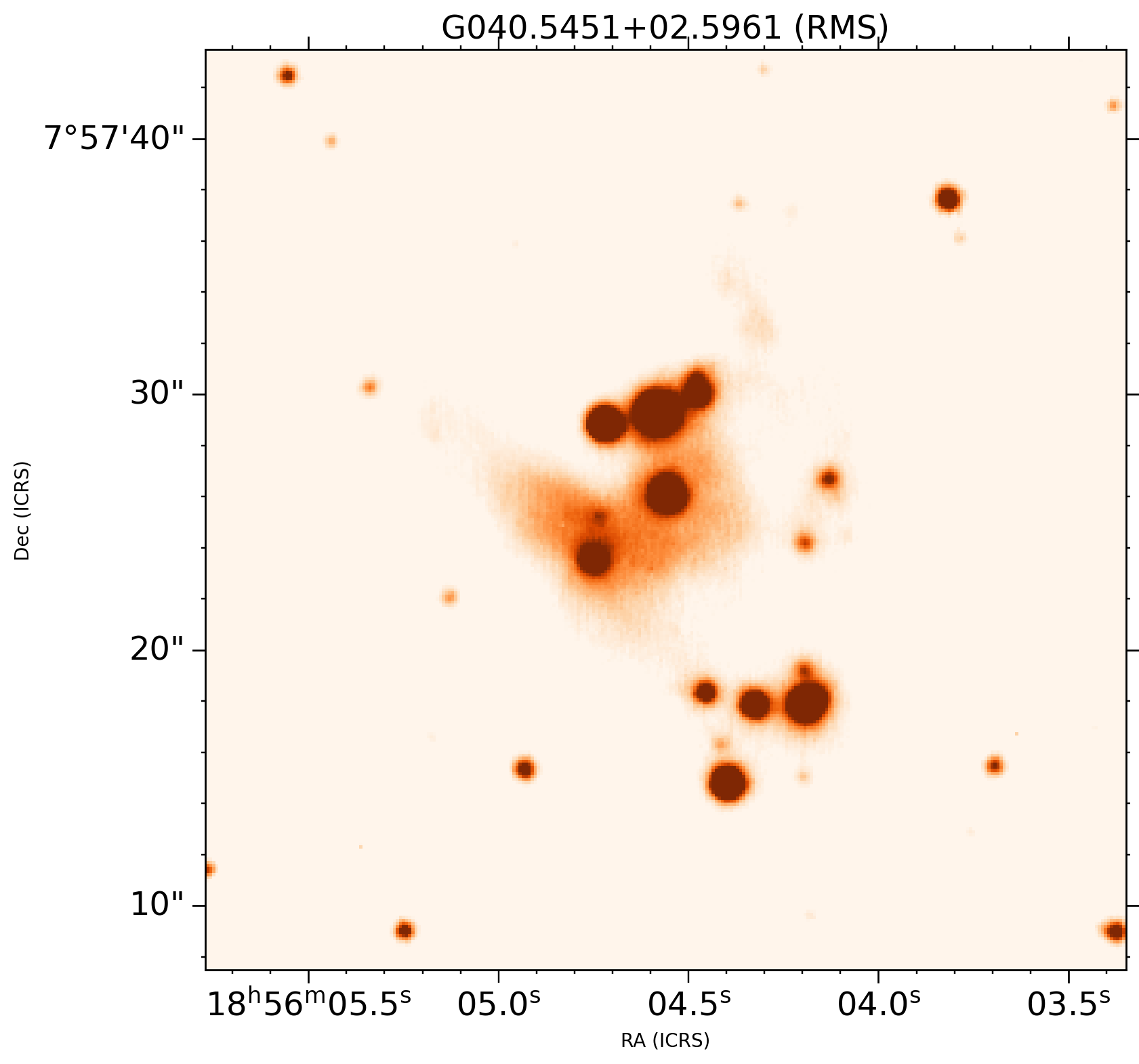}
            \label{fig:G040_RMS}
        \end{subfigure}
        \caption[A comparison between 2MASS (top), UKIDSS (middle) and RMS (bottom) infrared K-band images for the YSO G040.5451+02.5961. The superior resolution of the RMS and UKIDSS/VVV images allows for the detection of companions which were previously unresolved in 2MASS.]
        {\small A comparison between 2MASS (top), UKIDSS (middle) and RMS (bottom) infrared K-band images for the YSO G040.5451+02.5961. The superior resolution of the RMS and UKIDSS/VVV images allows for the detection of companions which were previously unresolved in 2MASS.} 
        \label{fig:G040_comp}
\end{figure}

UKIDSS/VVV are able to resolve objects almost as well as the RMS images, due to their similar resolution, but have the added benefits of multi-colour information and a deeper limiting magnitude, similar to the lower-resolution 2MASS survey. These differences are visible in \autoref{fig:G040_comp}, where the four resolved bright objects in the centre of the RMS and UKIDSS images appear as a single luminous object in the 2MASS image.

\section{Source Detection}
The RMS $K-$band images did not have a pre-existing point source catalogue and this was constructed using source detection code, while the UKIDSS and VVV surveys have point source catalogues readily available. The point source catalogues were tested against both the UKIDSS survey's own imaging and the RMS images, to determine the reliability of the catalogued sources. From the tested objects, there were no significant omissions or erroneous entries in the catalogue that could not be filtered out using flags or by simple visual inspection.

\subsection{Point source catalogues}
A region of 1.5 arcminute radius (to cover the same FoV of the RMS images) around each YSO was retrieved from the WFCAM Science Archive (\url{http://wsa.roe.ac.uk}) or the VISTA Science Archive (\url{http://vsa.roe.ac.uk}), depending on whether it was in the Northern or Southern sky respectively. The RMS coordinates were cross-matched with the catalogue data of the regions corresponding to each primary. The closest target to the inputted coordinates was initially assumed to be the primary, and a manual check was done for objects which had a significant separation between the coordinates of the RMS target and the UKIDSS target. Any unrelated point source which had been interpreted as the primary YSO was manually corrected.

One issue with the point source catalogues was the existence of duplicated and/or saturated sources. Objects brighter than $K$=11-12 could potentially be saturated, with some exhibiting ring-like artifacts which then are registered as multiple detections around the ring. Also some non-saturated point sources are entered more than once in the UKIDSS point source catalogue, even in the final merged source table. To overcome this, the UKIDSS and VVV catalogues have additional flags that filter out objects with quality control issues (objects with kppErrBits $<$ 256 were kept). Visual inspections had to be done afterwards to manually remove some outlying sources and ensure no false detections were still included. 2MASS photometry was used in place of UKIDSS/VVV for saturated sources; when the 2MASS magnitude was brighter than the UKIDSS/VVV saturation limit, the 2MASS magnitude was used.

To detect objects in the RMS images, a point source catalogue was constructed using the source detection program DAOphot \citep{STE87} along with Astropy \citep{ASTPY1,ASTPY2}. Objects with a brightness $3\sigma$ above the image's background value were classed as true detections. DAOphot also provides estimates for the magnitude of each source along with its uncertainty, which were calibrated using UKIDSS K-band photometry.

\subsection{Completeness}
To determine the completeness of the data, the limiting magnitude of the RMS images was determined by creating fake sources in the images. Multiple artificial Gaussian sources of varying intensity and distance from the parent object were injected into the images, using Astropy's Gaussian2DKernel function \citep{ASTPY1}. For each image, four copies were created which then had $\sim$10 artificial sources injected into them; the results of the analysis for each copy were compiled together into a single data set for each image. These artificial sources were set to the same FWHM as the average seeing of the sources in the images. The minimum intensity at which the artificial sources would be detected by DAOphot would correspond to the limiting magnitude of the images; the distance was also varied to see how closeness to the central MYSO would affect this limit. A hindrance of detecting faint close-in companions will affect the accuracy of the companion statistics. 
It was concluded that in the RMS images, close-in binaries at distances within $\sim$1.5 arcsec of the primary would not be consistently detected, and the limiting magnitude in these inner regions can be up to $\sim$3 magnitudes brighter than at larger separations. This is due to extended emission or crowded regions leading to source confusion or obfuscation. At $\sim$2 arcsec and beyond, the sensitivity improves and stars around 3.5 mag fainter than the primary are detected. Artificial star analysis was also performed on the UKIDSS/VVV images to show the difference in the detection ability of DAOphot for each survey. UKIDSS/VVV struggle more within 2 arcsec of the primary but perform similarly to the RMS images beyond that.

These comparisons demonstrated the benefits and caveats of each of these surveys: the RMS, UKIDSS and VVV surveys can probe deeper than the NaCo images used in \cite{POM19}, allowing fainter objects to be detected. However, the NaCo images have a much better resolution meaning that objects within 1 arcsecond of the primary (or other nearby objects) may not be resolved in the RMS/UKIDSS/VVV survey data. The RMS image data takes the middle ground, having a better resolution than UKIDSS/VVV but worse than NaCo, and a slightly worse limiting magnitude than UKIDSS/VVV. UKIDSS/VVV have the added benefit of full $J$-, $H$- and $K$-band photometry, providing more information on the companion candidates.

\subsection{Physical binary probability}
\label{prob}
An important factor to take into account is the fact that any detected potential companion may simply be a chance projection on the sky, and not be a physical binary companion. For each primary YSO, the density of background objects $\rho$ within 1.5 arcminutes was assessed to quantify how many objects are present in the nearby line of sight. This was done by sorting every background object in the region by its $K-$band magnitude, and then determining the number of background objects brighter than the putative companion by the total observed area in arcseconds$^2$. This allows us to assign a background source density to each background object which effectively scales with the brightness of the object in question; a bright object amongst more numerous fainter sources is more likely to be a companion than a faint source among equally faint background sources. Therefore the likelihood of an object being a physical companion has three dependencies in total: a) the separation of the object from the primary, b) the brightness of the object with respect to background sources, and c) the stellar background density. The further an object is from the primary, the fainter an object, or the denser the stellar background, the more likely an object is deemed a chance projection.

The Poisson distribution (\citealt{Albada1968, COR06}, see also \citealt{Halbwachs1988}) defines this probability:

\begin{equation}
	P=1-e^{-\pi d^2 \rho}
	\label{eq1}
\end{equation}
where $d$ is the distance from the primary to the potential companion in arcseconds and $\rho$ is the background density of objects brighter than the potential companion in arcsec\textsuperscript{-2}. The full 1.5 arcminute radius of the retrieved catalogue data was used to determine the background density. Spot checks were performed to ensure that the chance projection probability of objects scaled correctly with each of the different dependencies.

\subsection{Physical companions}

For each primary in the sample, objects in their neighbourhood were investigated to see if they could be classed as probable companions. The probability of each candidate being a visual binary was calculated using \autoref{eq1}, and those with $P_{chance}>20$ per cent were disregarded as probable chance projections. The multiplicity and companion fractions (MF and CF) were calculated for the potential companions detected within this limit, defined by the formulae: MF $= \frac{N_{\mathrm{m}}}{N_{\mathrm{tot}}}$ and CF = $\frac{N_\mathrm{b} + 2N_\mathrm{t} + 3N_\mathrm{q} + ...}{N_\mathrm{s} + N_\mathrm{b} + N_\mathrm{t} + N_\mathrm{q} + ...} $, where $N_\mathrm{m}$ is the number of multiple systems, $N_{\mathrm{tot}}$ is the total number of systems, $N_\mathrm{s}$ is the number of single systems, $N_\mathrm{b}$ is the number of binary systems, $N_\mathrm{t}$ is the number of triple systems, $N_\mathrm{q}$ is the number of quadruple systems, and so on.

\begin{figure}
        \centering
        \begin{subfigure}[t]{0.475\textwidth} 
            \label{fig:dmag10}
            \includegraphics[width=\columnwidth,left]{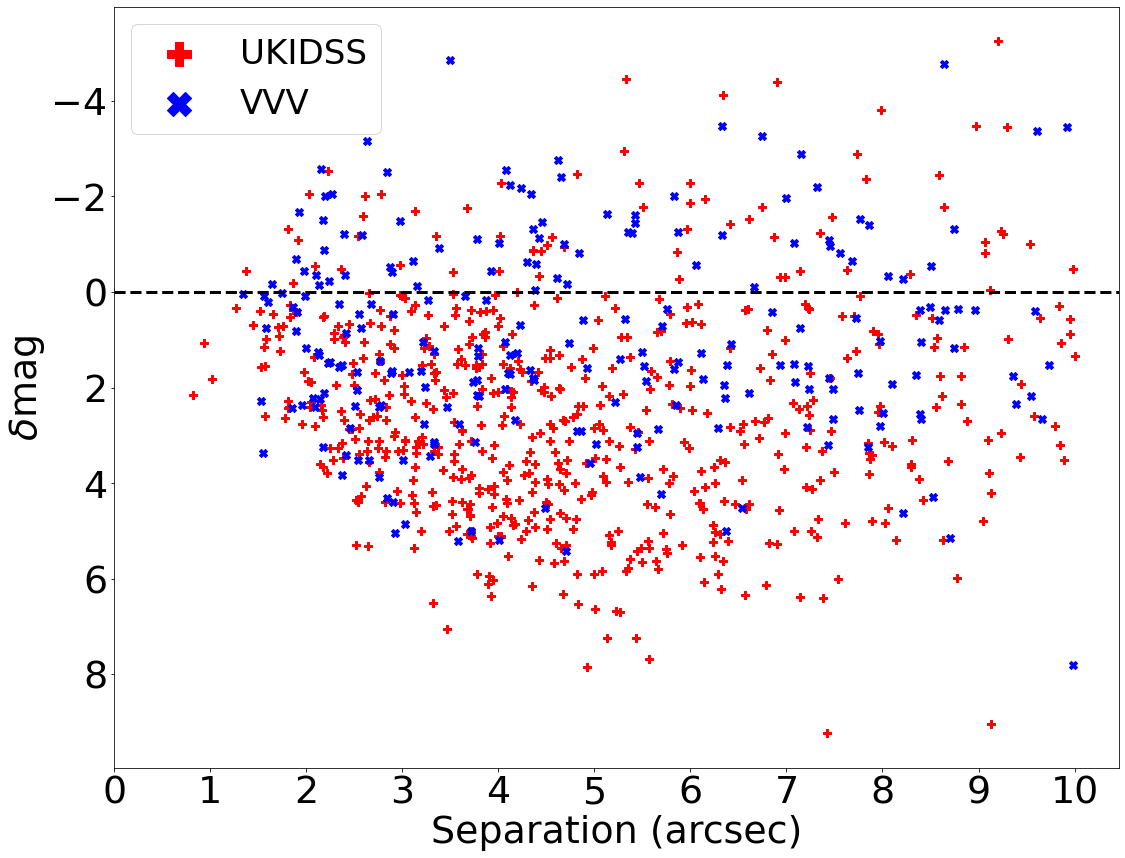}
            \vspace*{2mm}
        \end{subfigure}
        \begin{subfigure}[t]{0.49\textwidth} 
            \label{fig:dmag30}
            \includegraphics[width=\columnwidth,left]{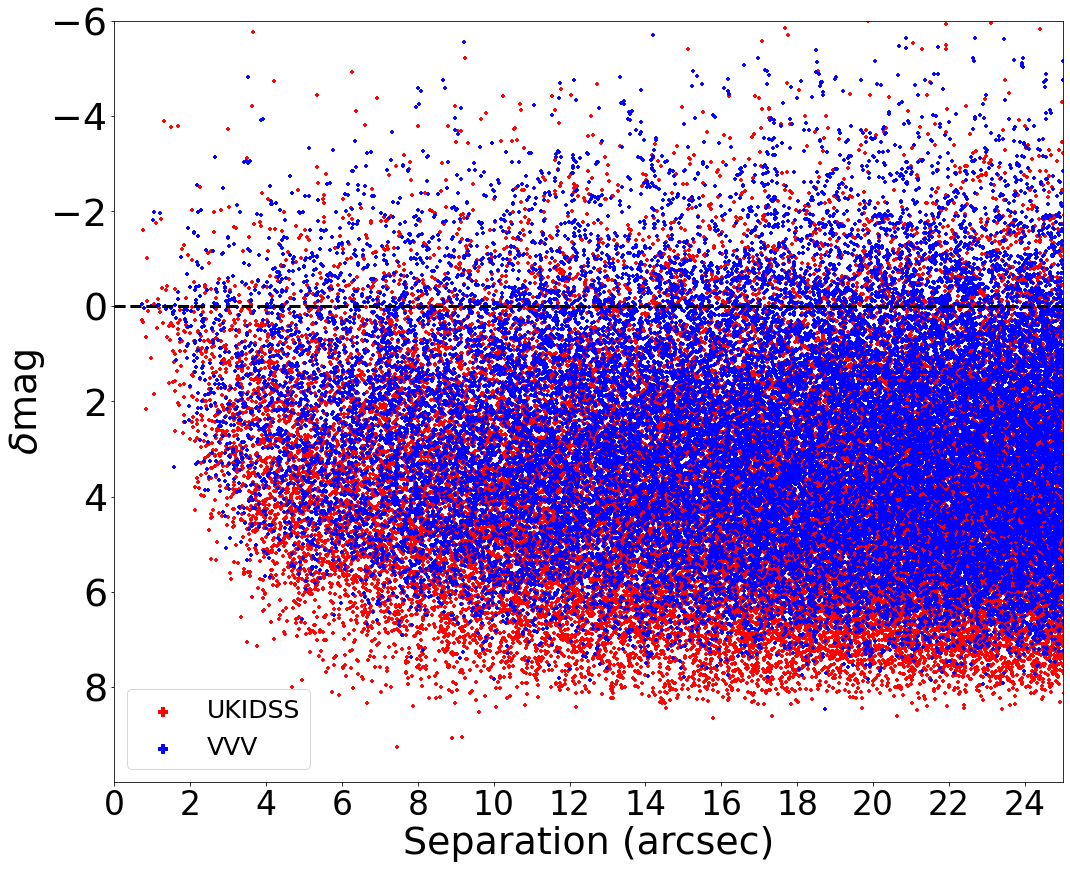}
            \vspace*{2mm}
        \end{subfigure}
        \caption{\small Top: The difference in the $K-$band magnitude between the companions with $P_{chance}<20$ per cent and their primary in arcseconds plotted against the separation. The UKIDSS companions are shown with red crosses while the VVV companions are shown with blue pluses. It is  apparent that very few VVV objects have a $\delta$mag greater than 3, while numerous UKIDSS objects have $\delta$mag up to and greater than 6. Objects with $\delta$mag$<0$ are brighter than the primary, this is presumably because of the smaller extinction to these sources. Bottom: the same plot but now including all objects up to 25 arcseconds.} 
        \label{fig:dmag}
\end{figure}

\autoref{fig:dmag} illustrates which objects are classified as binary companions and which are not. It shows how companion brightness relative to the primary ($\delta$mag) relates to proximity to the primary. A clear dearth of fainter detected sources is visible at $<$2 arcseconds, demonstrating that only the brightest objects can be detected at very close separations. Additionally, there seems to be a binary "sweet spot" with more companions between 3-6 arcseconds, and a drop-off at $>$7 arcsec. This drop-off can be understood when exploring \autoref{eq1}, as a fainter object at a large separation is unlikely to be registered as a probable binary companion at all. It therefore makes sense that companions of any brightness are more likely to be found at a mid-point, such as this "sweet spot". 9 arcseconds was the chosen upper limit for companion detection because there is a distinct flattening in the number of objects in the field beyond this point in each of the samples; this is where the random distribution of background stars is  probed. We note that quite a few companions are apparently brighter than the primary object (which for the purposes of this paper is the MYSO). This can be explained by the fact that the extinction towards the MYSO is often dominated by its circumstellar material \citep[e.g.][]{FRO19,FRO21}. In certain instances, it will then be fainter than its nearby companions.  

The multiplicity fractions for each sample can be found in \autoref{tab:mfs}. We will investigate this further in Section 4.2.

\begin{table}
	\centering
	\caption{Multiplicity results for each sample, separated into subsets based on YSO mass. Objects with M$>$8M$_\odot$ are classed as high-mass.}
	\label{tab:mfs}
	\begin{tabular}{lllllll}
		\hline
		Sample & Subset & MF & CF\\
        & & (\%) & (\%)\\
		\hline
		UKIDSS & All & $65\pm 4$ & $147\pm 6$\\
		& High-mass & $67\pm 5$ &\\
		& Low-mass & $66\pm 6$ &\\\\

        VVV & All & $53\pm 4$ & $84\pm 5$\\
		& High-mass & $54\pm 8$ &\\
		& Low-mass & $54\pm 5$ &\\\\

        UKIRT/RMS & All & $64\pm 8$ & $139\pm 9$\\
		& High-mass & $60\pm 11$ &\\
		& Low-mass & $69\pm 14$ &\\
		\hline
	\end{tabular}
\label{log}
\end{table}

\subsection{Mass ratios}

Here we make an attempt at deriving the mass ratios of the systems that were detected above. Given that much more information is available for the primary YSO objects than for the secondary stars, the way we determine their masses is different. The masses of the primary YSO objects are determined
using the bolometric luminosities listed in the RMS catalogue and the mass-luminosity relations from \cite{DAV11}. Although these are based on the main sequence, they are representative of the pre-main sequence masses as the stars evolve fairly horizontally on their pre-main sequence evolutionary tracks \citep{Bressan2012}.

We cannot apply this method to the companions that are reported here as only near-infrared photometry is available for them.  However, under the assumption of a main-sequence nature of the objects, we can use the absolute $J$ or $K$-band magnitudes as a proxy for the mass. This was for example done for B and Be stars using $K$-band photometry by \citet{OUD10} who derived: 

\begin{equation}
	log(M/{M_\odot})=-0.18K_{\mathrm{abs}}+0.64,
	\label{eq2}
\end{equation}
where $K_{\mathrm{abs}}$ is the extinction-corrected absolute $K$-band magnitude.
Following \citet{OUD10}, the derivation of \autoref{eq2} was performed for the $J$-band, and so the $J-$band magnitude can also be used as a proxy for the mass:

\begin{equation}
	log(M/{M_\odot})=-0.16J_{\mathrm{abs}}+0.65,
	\label{eq3}
\end{equation}
where $J_{\mathrm{abs}}$ is the extinction-corrected absolute magnitude. As we do not know whether the companion objects have infrared excess emission due to circumstellar dust, the $J$-band would be preferable as hot dust is more prevalent at the $K$-band. 

The challenge is to determine the extinction towards the objects. In \cite{POM19}, who only had $K$-band photometry available, lower and upper limits to the extinction (and by implication companion masses and mass ratios) were determined using  the foreground extinction and the  `total' extinction (foreground + circumstellar extinction) of the {\it primary} object respectively. The former was sourced from extinction maps and the latter using the observed $JHK$ colours of the primary as per \citet{COO13}. 

Here we can take this a step further as the multi-colour information available in the UKIDSS/VVV point source catalogues allows estimations of the total extinction of the companion itself rather than that towards the primary, whose own circumstellar extinction is likely to be larger owing to their embedded nature.  Below we make estimates of the foreground extinction from extinction maps and of the total extinction using the near-infrared colours of the objects.

The dust map chosen for our foreground extinction estimates was Bayestar19 \citep{BAYESTAR19}, a three-dimensional map of dust reddening across most of the Galaxy. However, Bayestar19 does not cover the Southern sky at $\delta<-30$. For these objects, we chose to use the dust maps of Stilism \citep{STILISM17}, which cover the whole Galactic plane but have a lower distance cutoff than Bayestar19. Therefore Bayestar19 was used as our main dust map, while Stilism was used for the regions that Bayestar19 does not cover. As a result of this caveat of Stilism, mass ratios of the more distant objects in the Southern sky derived using foreground extinction may be  less reliable.

To determine the total extinction towards a companion, $J-H$ photometry from UKIDSS/VVV was used to estimate $A_V$ as in \cite{COO13}, where the photometry was compared to the expected colours of a MS B0 star. Not every YSO in UKIDSS and VVV has $J$-band photometry; where $J-H$ photometry was unavailable, $H-K$ was used instead. Objects which exhibited a `negative extinction' due to their UKIDSS/VVV colours had their extinction set to zero. Once the companion's $K$-band photometry was corrected for extinction, the distance to the primary was used to convert the apparent magnitudes into absolute ($K_{\mathrm{abs}}$) magnitude; the distances were retrieved from the RMS catalogue. 

Two sources for the mass proxy were used as each have their own caveats: using $K-$band photometry can result in a mass overestimation due to dust excess, while $J-$band photometry may lead to an underestimate as a result of increased scattering. Using the primary mass determined from the RMS luminosities, estimates of the mass ratios could then be calculated. 

The total extinction towards the primaries was not used; instead, the extinction values of the companions themselves are used  as they provide a more accurate correction for the $K-$ and $J-$band magnitudes of the companions, especially ones at larger separations which are unlikely to share the same extinction as their primary. 

For the purposes of this paper, we define the mass ratio as $q = M_{\mathrm{comp}}/M_{\mathrm{prim}}$, where $M_{\mathrm{comp}}$ is the mass of the companion and $M_{\mathrm{prim}}$ is the mass of the primary. We will discuss the results in the next section. 

\begin{figure}
    \centering
    \begin{subfigure}[t]{0.475\textwidth}   
        \centering
        \includegraphics[width=\columnwidth]{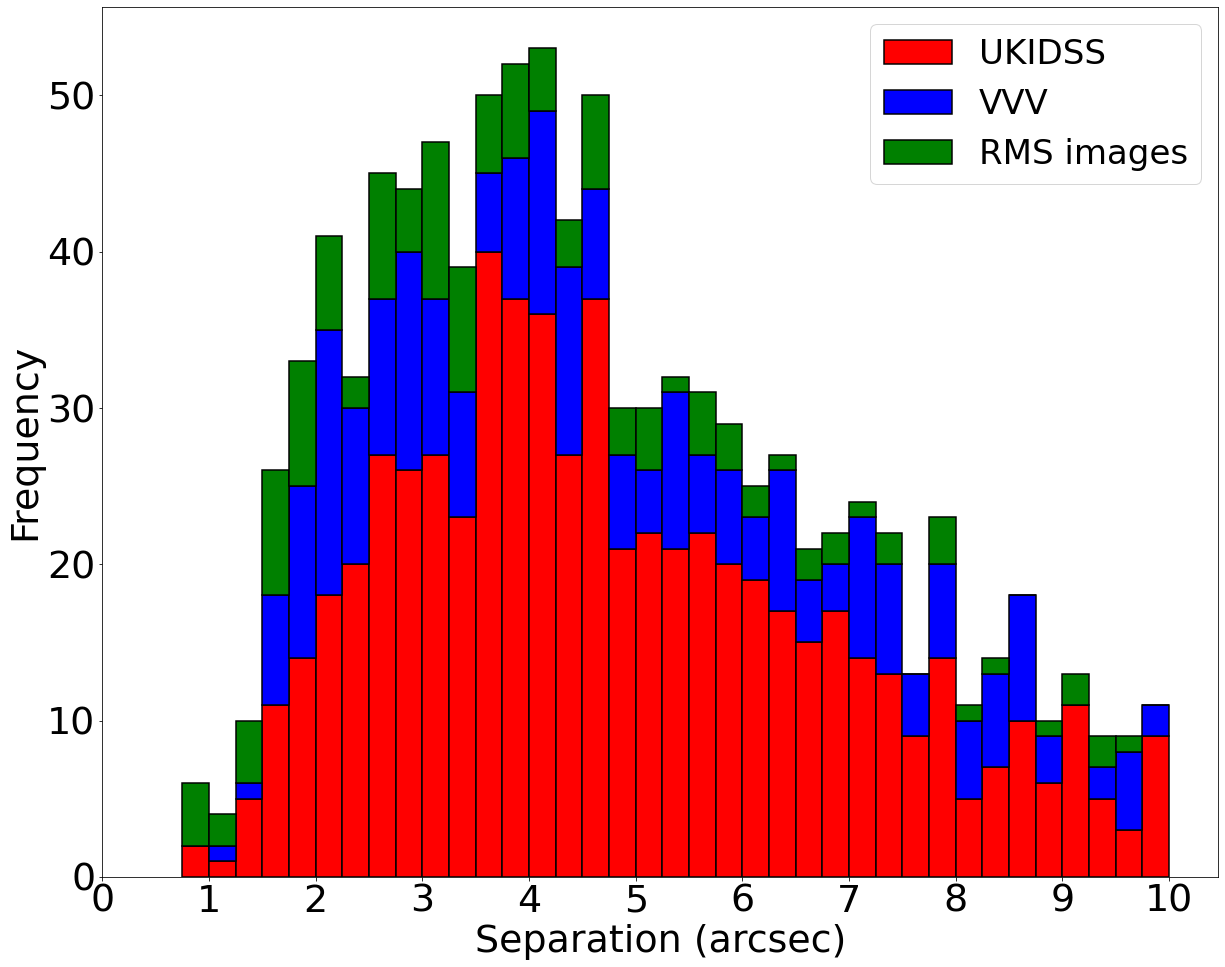}
        \label{fig:comp_sep_hist}
        \vspace*{2mm}
    \end{subfigure}
    \begin{subfigure}[t]{0.475\textwidth}   
        \centering
        \includegraphics[width=\columnwidth]{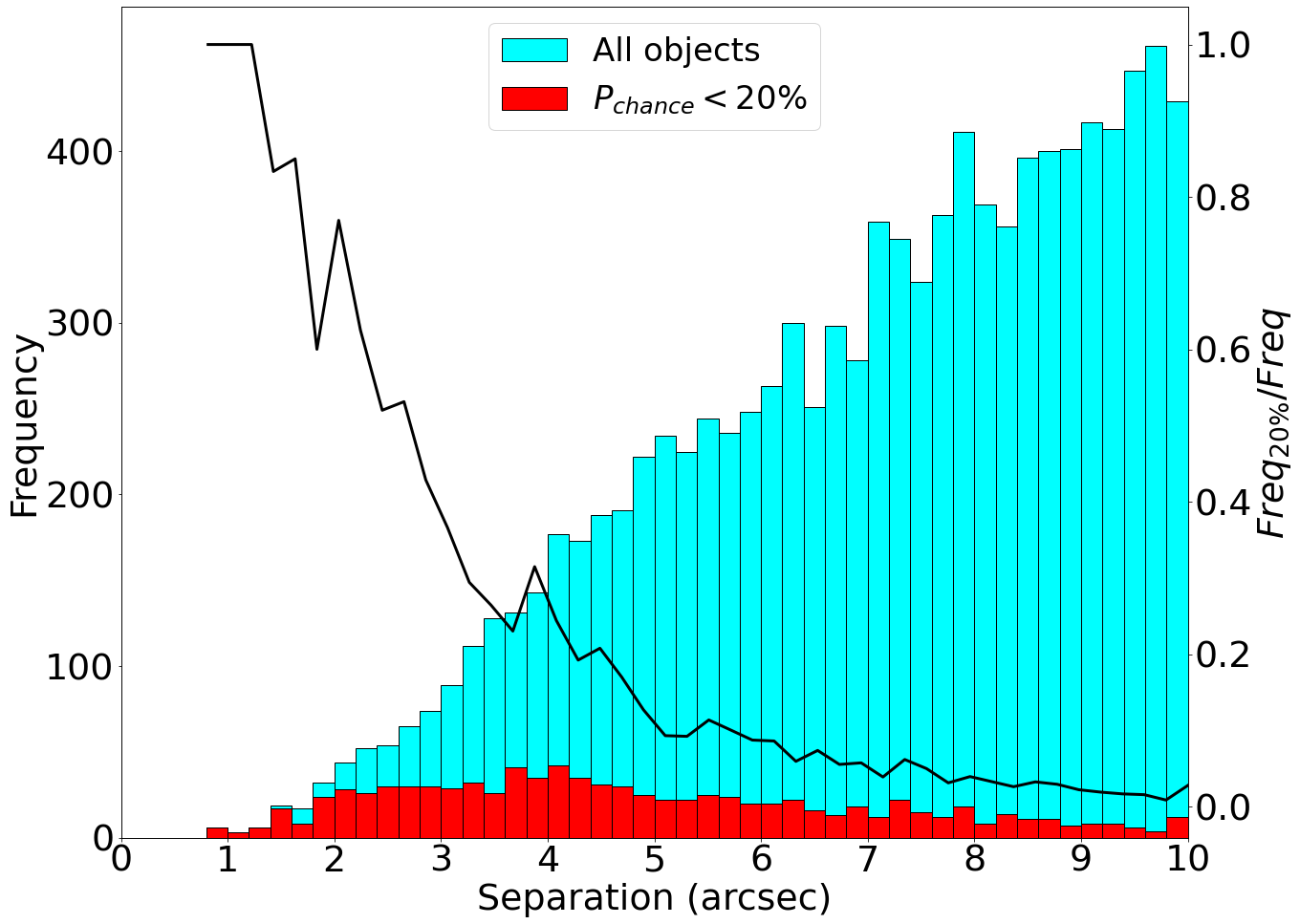}
        \label{fig:comp_all}
    \end{subfigure}
    \caption{\small Top:  histogram of the angular separation between the detected companions and their primaries, colour-coded for the three samples. Bottom: histogram of the separation between all detected companion candidates and their primaries. The red objects have a $P_{chance}$ less than 20 per cent of being a background star. The black line represents the ratio between the frequency of $P_{chance}<20$ per cent objects and the whole sample at each separation.}
    \label{fig:sepplots}
\end{figure}

\begin{figure*}
    \centering
    \includegraphics[width=0.95\textwidth]{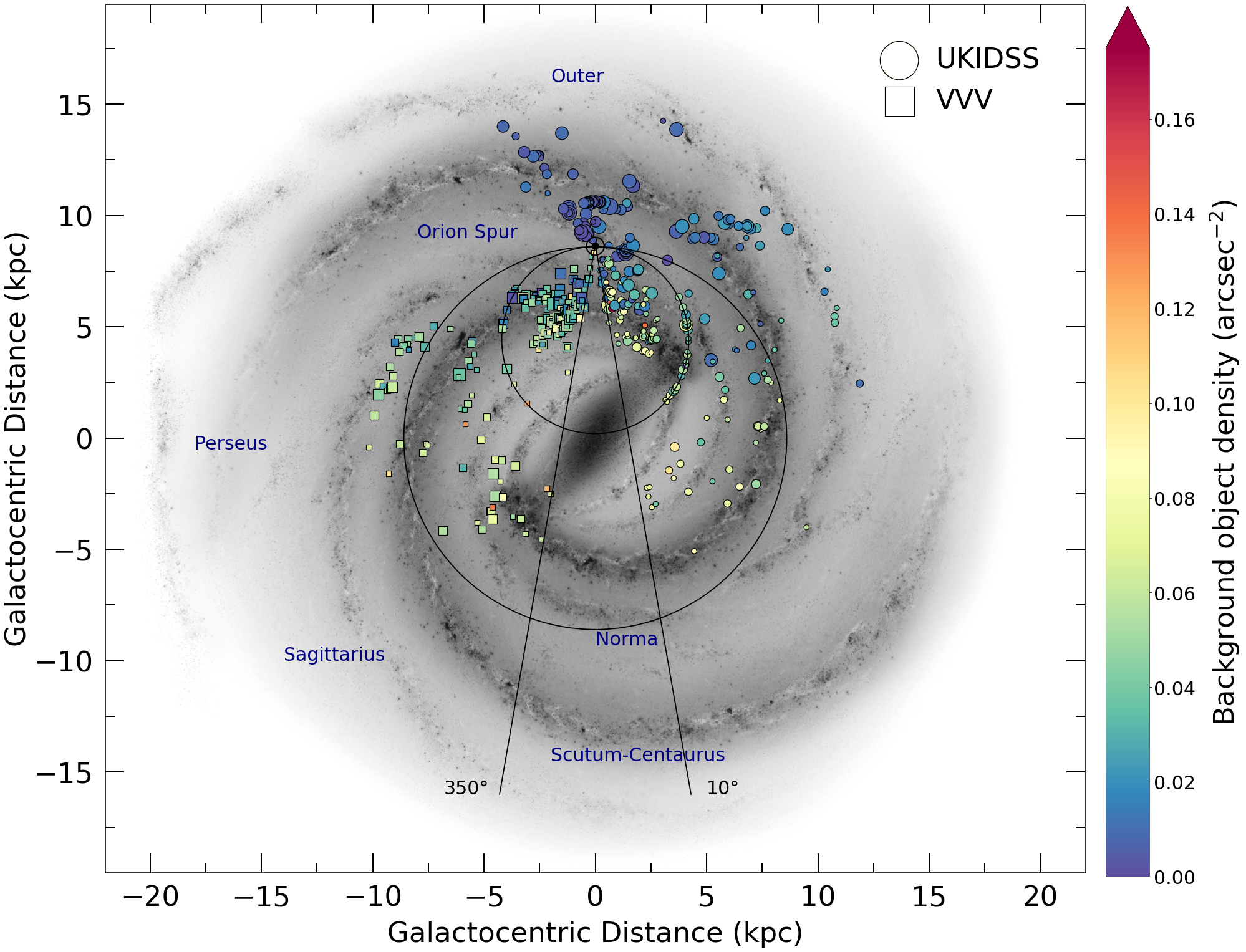}
    \caption{\small A diagram of the Galactic plane showing the position of the YSOs in our UKIDSS (circle) and VVV (square) samples. The larger ring represents the Solar circle and also shows the divide between the `inner' and `outer' Galaxy. The points are coloured based on the YSO's surrounding background object density, and also are sized depending on the number of detected companions; larger points are YSOs with more companions. The binary fraction in the outer Galaxy appears larger than in the inner Galaxy. This is likely due to the lower stellar background count increasing the companion probability in Eq. 1. The two black lines enclose the Galactic centre region which was not included in the RMS survey due to confusion regarding the sources and their distances.} 
    \label{fig:galaxy_pos}
\end{figure*}

\section{Results \& Discussion}
The tables of all detected companions in UKIDSS/VVV and the RMS images can be found in \autoref{appB} and \autoref{appC} respectively. The distribution of companion separations can be seen in \autoref{fig:sepplots}. The top plot shows the separation distribution of each sample out to 10 arcseconds. The bottom plot includes all objects in the field. We measure three YSO binary fractions, one for each sample: for UKIDSS the MF = $65\pm 4$ per cent; for VVV, it is $53\pm 4$ per cent; and for for the UKIRT image sample, the MF = $64\pm 8$ per cent. In each sample, a significant fraction of the mass ratios were greater than 0.5. 

\subsection{Statistical differences in surveys and galactic regions}
\label{statdiffs}

The existence of different companion statistics of the UKIDSS and VVV surveys is counter-intuitive as the surveys are highly comparable. The solution to this conundrum may be found in the fact that UKIDSS probes not only the inner Galaxy, like VVV, but also the outer Galaxy: the outer section of the Galactic plane, surveyed by UKIDSS, has a binary fraction of 80$^{+6}_{-7}$ per cent. The Northern inner part of the Galaxy, also surveyed by UKIDSS, has a binary fraction of 54$\pm$6 per cent, much lower than the outer galaxy. The UKIDSS inner region aligns statistically with the VVV fraction of 53 per cent, which only surveys the Southern inner galaxy. \autoref{fig:galaxy_pos} shows the different regions of the Galaxy and the surveys that probed them.

Hence, at first sight it would appear that the  multiplicity of MYSOs is larger in the outer Galaxy than in the inner Galaxy. Given that the metallicity of stars in the Galaxy decreases with Galactocentric radius (e.g. \citealt{Mendez2022}), and that the close binary fraction of Sun-like stars increases with decreasing metallicity \citep{Badenes2018,Moe2019}, it would be tempting to assume that the higher binary fraction we observe in the outer regions is due to the stars having lower metallicities. However, the metallicity dependence is only observed for close, Sun-like, binaries, which can be explained in terms of more efficient fragmentation in low metallicity environments \citep{Bate2019}, while we clearly deal with wider and more massive binaries in this paper. Instead, we note that \autoref{eq1} which computes the probability of a source to be a physical companion  has a built-in dependence on the stellar background density.  Indeed, when considering the inner Galaxy, the stellar background appears more dense (as indicated in the colour table in \autoref{fig:galaxy_pos}), and so according to \autoref{eq1} the probability for any companion to be a background source will be larger than in a lower density background such as in the outer Galaxy where the stellar background appears to be less dense. Thus, the likelihood of nearby objects meeting the criteria of a physical companion in low background density regions is increased, driving up the observed multiplicity fraction in the outer Galaxy.

This also shows that the outer, less dense region of the Galaxy as surveyed by UKIDSS is responsible for the significantly larger binary fraction in UKIDSS compared to VVV, and that a large number of objects are missed in the inner Galaxy due to observational bias. The outer galaxy fraction of 80$^{+6}_{-7}$ per cent suggests a very high multiplicity in YSOs, approaching 100 per cent as already inferred by \citet{POM19} based on general arguments. 

\subsection{Multiplicity statistics}
Despite the similar limiting magnitudes and resolutions between UKIDSS and VVV, the MF and CF of the VVV sample are significantly lower than that of the UKIDSS sample (and the RMS imaging sample). As mentioned above, this is due to differences in survey background density. When accounting for this by only including the `inner' region of UKIDSS with similar average background density to VVV, the multiplicity fractions of the two samples are within agreement, showing uniformity between the two samples. Across the UKIDSS and VVV surveys, the detected companions have a mean angular separation of 4.8", with a minimum of 0.8", a maximum of 9" and a standard deviation of 1.9". The companions have a mean physical separation of 17900 au, ranging from 910-121,000 au with a spread of 15500 au.

126 companions were found using the RMS images, with 106 of them associated with YSOs covered by the UKIDSS survey. From these 106 companions, 61 (58 per cent) were also detected in UKIDSS. The companions detected in both samples were generally the furthest from their primaries; closer companions can be detected thanks to the higher resolution of the RMS images, while UKIDSS struggles in these close-in regions. Additionally, objects that were lacking UKIDSS photometry would not have been detected as a companion in UKIDSS. 

Although the MF of the RMS imaging sample is within the uncertainties of that of the VVV survey, the CF is significantly higher. This can once again be explained by the survey density discrepancies mentioned above leading to more companions being detected in the 'outer' regions.

\begin{figure}
    \centering
    \begin{subfigure}[t]{\columnwidth}
        \centering
        \includegraphics[width=\textwidth]{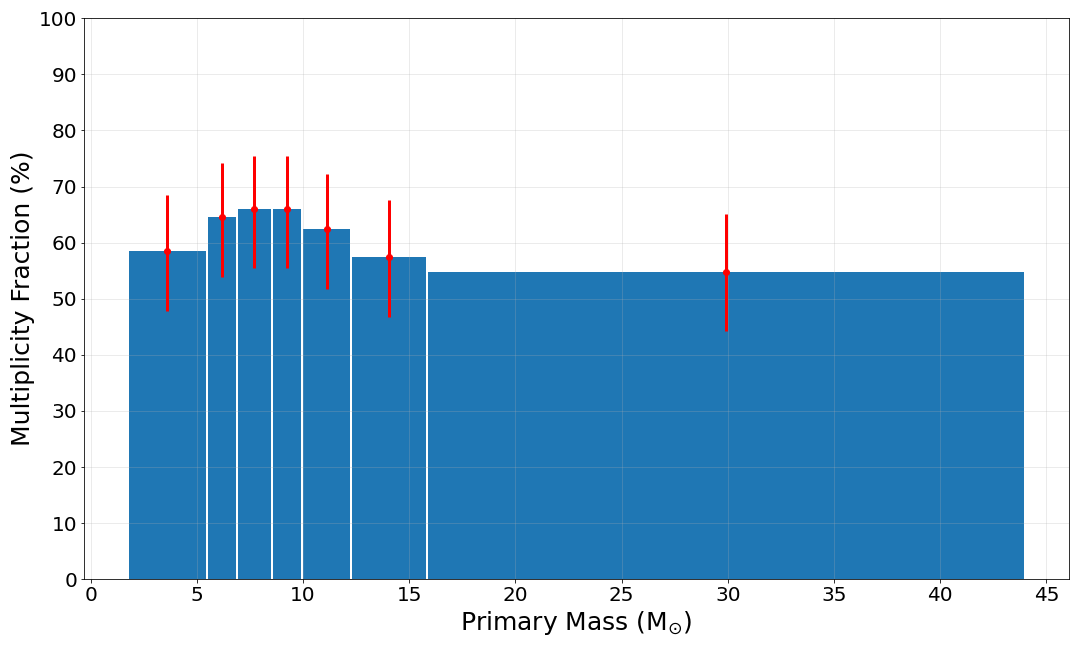}
    \end{subfigure}
    \hfill
    \begin{subfigure}[t]{\columnwidth}
        \centering
        \includegraphics[width=\textwidth]{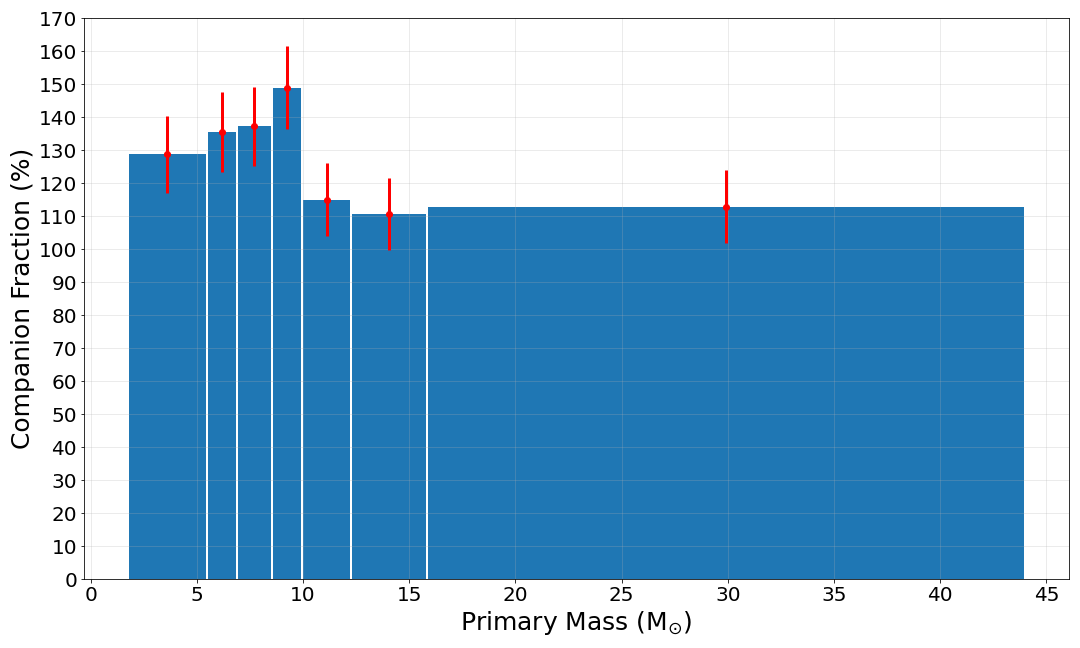}
    \end{subfigure}
    \caption{Top: The multiplicity fraction of different primary mass bins. Each bin contains an equal number of objects. The red error bars are derived from binomial confidence intervals. This shows a relatively flat distribution, and demonstrates that multiplicity generally is not affected by primary mass. Bottom: The companion fraction of different primary mass bins. Here there seems to be a hint of a drop-off in the number of companions formed per system around 10 $M_\odot$.}
    \label{fig:mass_MF_hist}
\end{figure}

The effect of primary mass on a YSO's multiplicity can be seen in \autoref{fig:mass_MF_hist}. It is clear that the primary YSO mass does not have a significant effect on whether the YSO forms at least one companion, save for a relatively small peak between 5-12 M$_{\odot}$ which can be accounted for by the uncertainty. Therefore it can be asserted that primary mass does not determine whether a companion is formed during the birth of a star. However it is apparent from the bottom plot of \autoref{fig:mass_MF_hist} that the frequency of companions per system exhibits a slight drop-off at $\sim$ 10 $M_\odot$; this is not enough of a drop-off to infer a significant feature.

\subsubsection{Comparison with previous MYSO surveys}
The fractions calculated for all three of our high-mass subsets are higher than that in \cite{POM19}, who report MF = $31\pm8$ per cent and CF = $53\pm9$ per cent for their sample of MYSOs. However this is due in part to the improved magnitude depth of our samples over the NaCo sample, meaning fainter companions not picked up by Pomohaci are more likely to be detected in the deeper IR surveys or our RMS images. Also the separations probed in each sample are different; the NaCo survey was able to probe closer to the primaries but it was only complete out to 3 arcseconds, as opposed to 9 arcseconds in our survey. By using the survey limits of \cite{POM19} with our survey, we can make a like-for-like comparison. A separation limit of 3 arcsec and a magnitude limit of 4.5mag fainter of the primary were used to match the two surveys, which gives us fractions of MF = $38\pm7$ per cent and CF = $48\pm7$ per cent, which are well within the uncertainties of the \cite{POM19} survey. The inner 0.6 arcsec of the Pomohaci sample contains no companions, which aligns with the fact that our closest detected companion is at 0.8". This suggests that there may be a dearth of close-in MYSO companions, however future work will probe the inner regions of MYSOs using spectroscopy to determine the true binary fraction at these separations.

A recent interferometric MYSO survey by \cite{KOU21} found a binary fraction of $17\pm15$ per cent in a sample of six MYSOs between $\sim$2 and 300 au, a lower fraction than reported in this work; however this uses a much smaller sample size while their separation range is also smaller. An MYSO multiplicity survey by Bordier et al. (2023, submitted) used $L'-$band imaging to study eight MYSOs at separations of 600-35,000 au, and found a multiplicity fraction of $62\pm13$ per cent, in close agreement with the statistics found in this paper. 

\subsubsection{Comparison with other massive star surveys}
Previous surveys have also investigated binarity in massive stars. \cite{SAN12} investigated the multiplicity of O and B main-sequence stars and found them to have a MF = 70 and 52 per cent, and CF = 130 and 100 per cent respectively for separations between 2-200 au. \cite{OUD10} found that a sample of B stars and a sample of Be stars had binary fractions of $29\pm8$ per cent and $30\pm8$ per cent respectively at separations between 20-1000 au. Looking at more recent surveys, \cite{BAN22} studied binarity in B-type stars in the young open cluster NGC 6231 and found a binary fraction of $52\pm8$ per cent when correcting for observational bias, agreeing with our MF. \cite{BOR22} reports a MF of 100 per cent from a sample of young O-stars within 120 au, which is much higher than our determined binary fraction but also probes much closer separations.

Direct comparison between these surveys is not an easy task due to a number of factors; the significant differences in separations probed, the observational conditions, sensitivities and techniques used, and the differences in evolutionary status. The resolution of the data used here means that the inner $\sim$1-1.5 arcsec of each YSO is essentially a blank spot, and so we are unable to probe regions in which other surveys have found varying levels of multiplicity.

To conclude, the multiplicity fraction of the YSOs investigated here agrees with previous MYSO multiplicity studies at similar separation ranges, and generally agrees with previous studies into the binarity of B stars.

\begin{figure}
    \centering
    \begin{subfigure}[t]{\columnwidth}
        \centering
        \includegraphics[width=\textwidth]{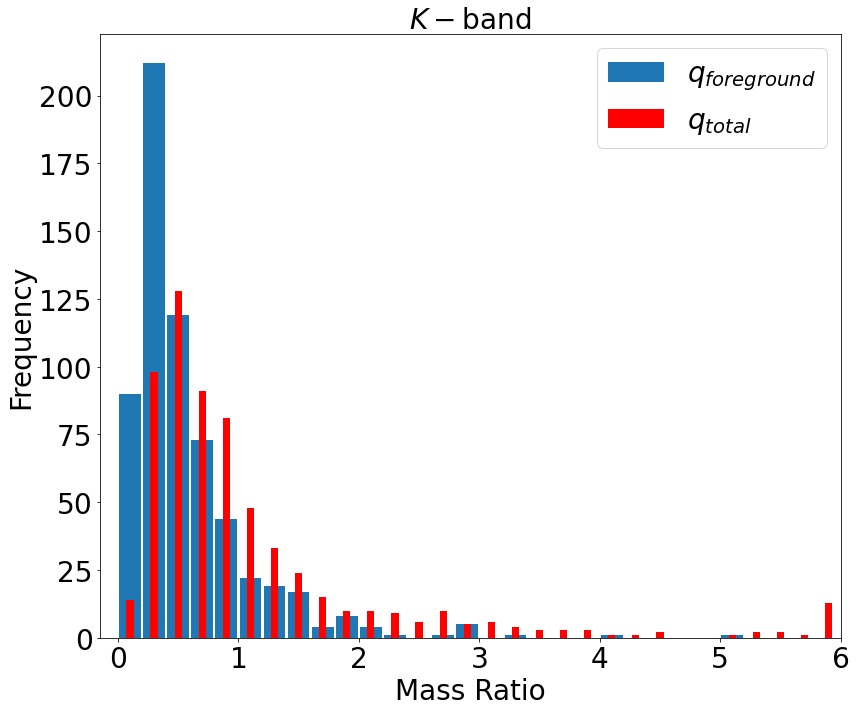}
    \end{subfigure}
    \hfill
    \begin{subfigure}[t]{\columnwidth}
        \centering
        \includegraphics[width=\textwidth]{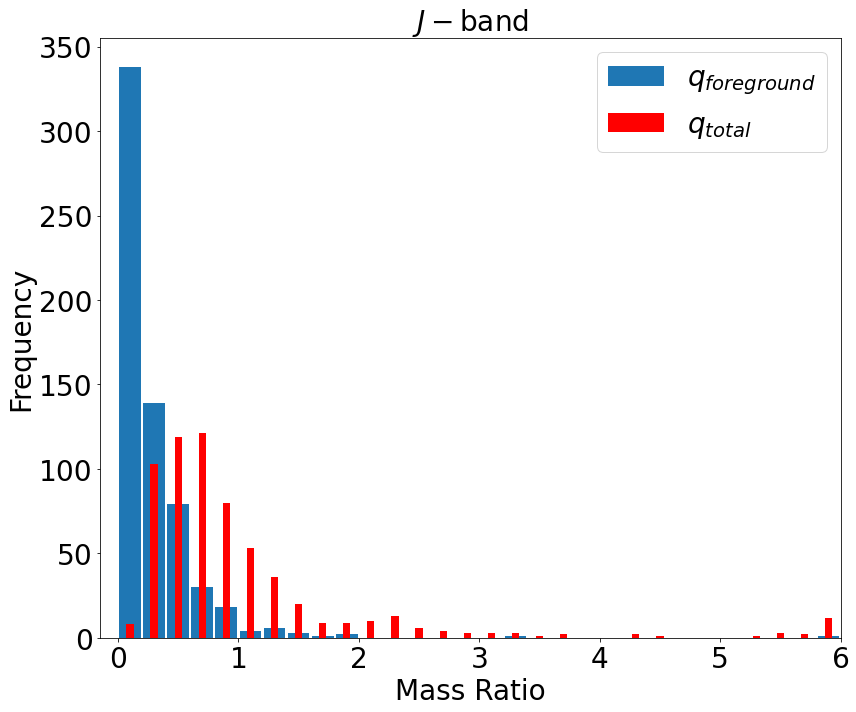}
    \end{subfigure}
    \caption{\small Histograms of the mass ratios of the detected companions with $P_{chance}<$ 20 per cent, using $K-$band (top) and $J-$band (bottom) mass proxies. The thick blue bars represent the mass ratios determined using only foreground extinction, while the thin red bars show the estimates using total extinction as derived from the near-infrared photometry. It can be seen that using the total extinction results in larger companion masses and ratios - see text for details. Any mass ratios greater than 6 are collected in the final bin.}
    \label{fig:massR_hist}
\end{figure}

\begin{figure*}
    \begin{subfigure}[t]{0.475\textwidth}
        \includegraphics[width=\textwidth]{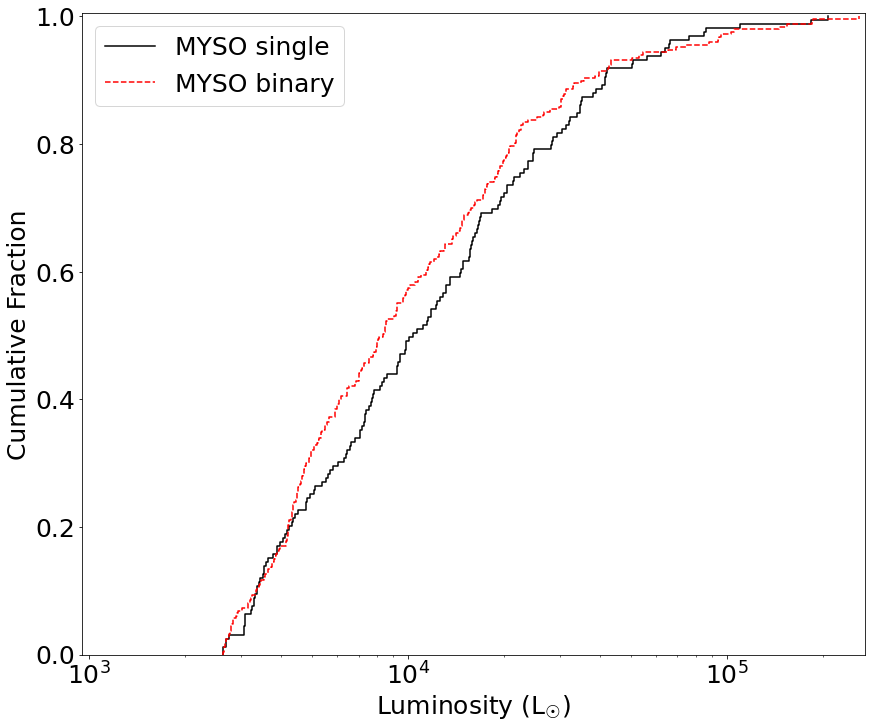}
    \end{subfigure}
    \hfill
    \begin{subfigure}[t]{0.475\textwidth}
        \includegraphics[width=\textwidth]{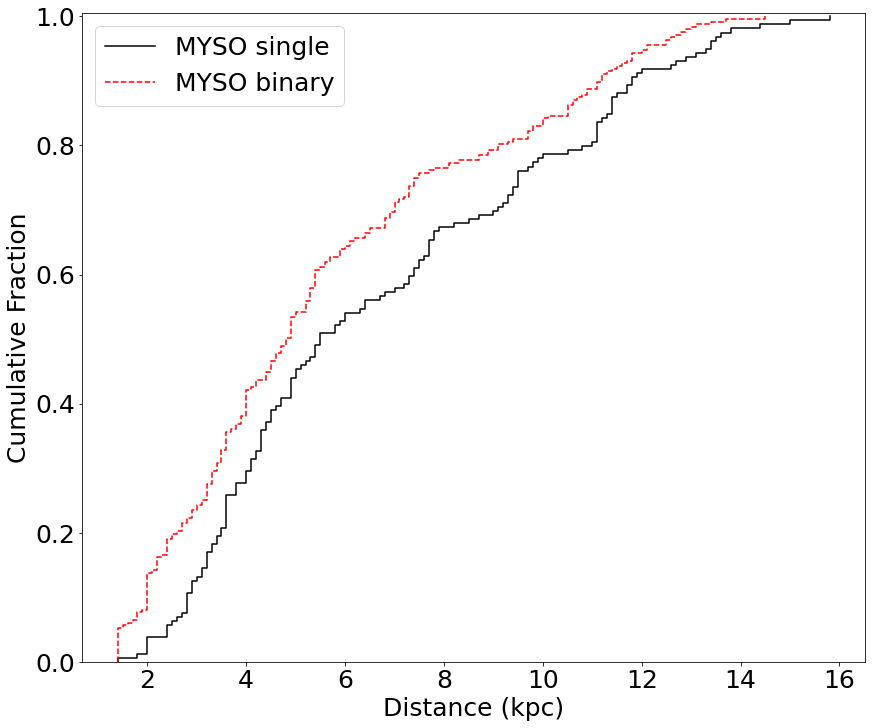}
    \end{subfigure}
    \caption{The cumulative distribution of luminosity (left) and distance (right) to the MYSO primaries. Single MYSOs are represented by the solid black line and binary MYSOs are represented by the dashed red line.}
    \label{fig:K-S_dist}
\end{figure*}

\begin{figure*}
    \begin{subfigure}[t]{0.475\textwidth}
        \includegraphics[width=\textwidth]{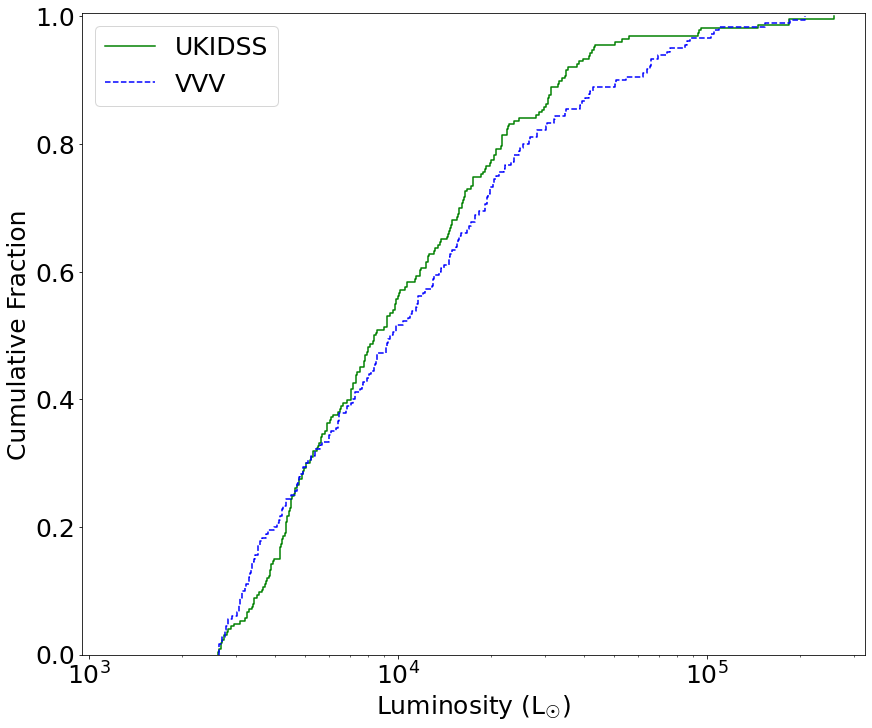}
    \end{subfigure}
    \hfill
    \begin{subfigure}[t]{0.475\textwidth}
        \includegraphics[width=\textwidth]{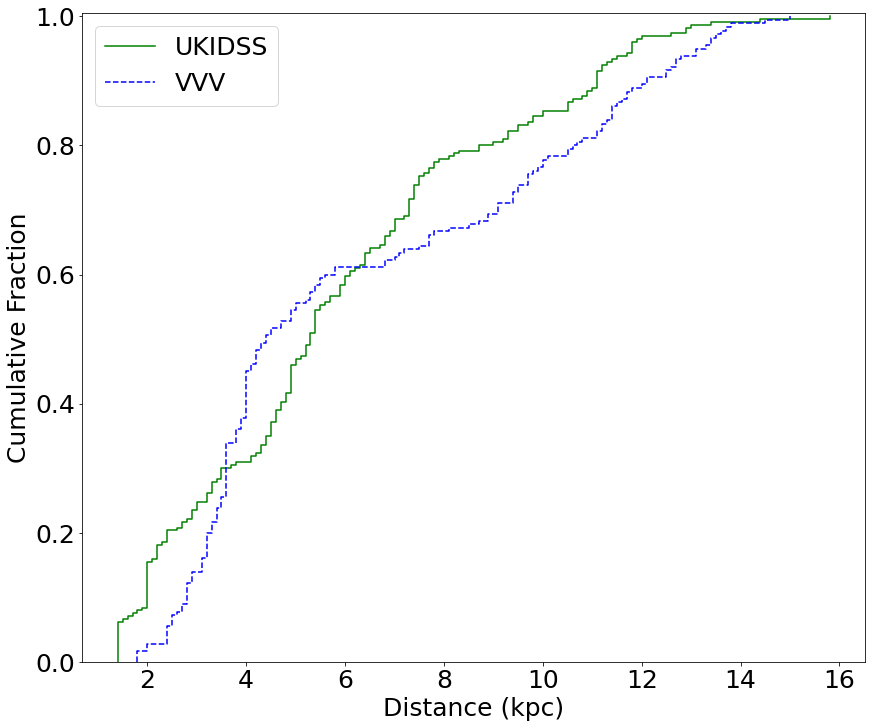}
    \end{subfigure}
    \caption{The cumulative distribution of luminosity (left) and distance (right) to the MYSO primaries. MYSOs found in UKIDSS are represented by the solid green line and MYSOs detected in VVV are represented by the dashed blue line.}
    \label{fig:K-S_ukidssvvv}
\end{figure*}

\subsection{Mass ratios}

Histograms of the mass ratio distribution resulting from the companions' mass estimates based on the $K-$band and $J-$band respectively can be found in \autoref{fig:massR_hist}. Using the estimation of foreground extinction ($A_{K,fg}$), the average mass of our companions is 5 $M_\odot$ and the average mass ratio is 0.5. Instead using the total extinction ($A_{K,tot}$), we find an average companion mass of 14 $M_\odot$ and an average mass ratio of 1.4. A significant fraction of companions have a mass ratio $q>0.5$. 

When using the $J$-band as proxy for companion masses, we find that the average companion masses and mass ratios are smaller than for the $K$-band estimates of both foreground extinction (3 $M_\odot$, $q=0.3$) and total extinction (12 $M_\odot$, $q=1.1$).

The masses determined for the companions are simple estimates from Eqs \ref{eq2} and \ref{eq3}, which assume the star is a MS star and that the  $J$- and $K-$band magnitudes are due to photospheric emission only. We find that the companions generally have large mass ratios ($>$0.5), especially from $A_{K,tot}$ estimates. Mass ratios significantly greater than 1 are likely due to excess emission due to hot circumstellar dust which leads to an overestimate of the the extinction and thus of the mass. Bearing these uncertainties in mind, the formal errors on our mass ratios are of order $\sim$20 per cent, mostly as a result of uncertainty in the determination of the bolometric flux of the objects \citep{MOT11}. Distance uncertainty is insignificant when taking the mass ratio as the same distance uncertainty applies to both the primary and secondary.

This sizable proportion of high mass ratios suggests an inconsistency with the binary capture formation scenario, which favours low mass ratios \citep{SAL55}. \cite{MOE17} found MS mass ratios consistent with random IMF sampling at large separations (similar to the separations probed here) but large mass ratios for close binaries. This also leads to a potential situation where the distribution of secondary separations in MYSOs may not be constant, and changes over time. Migration could be an explanation for this, as \cite{RAM21} suggest that stars may form in wide binary systems and migrate inwards over time to form tighter pairs.

More accurate estimates for extinction could be made using infrared excess determinations (e.g. through fitting spectral energy distributions) but this is outside the scope of this work for this very large sample size. 

\subsection{Are binary YSOs different from single YSOs?}
To see whether binarity has an effect on an MYSO, the samples were studied to look for differences in the properties of single MYSOs and MYSOs with one or more companions.

For single MYSOs, the average luminosity is 19,000 $L_{\sun}$ and the average distance is 6.7 kpc. The average luminosity of binary MYSOs is 18,000 $L_{\sun}$, with an average distance of 5.7 kpc. For comparison, the entire sample of UKIDSS/VVV MYSOs has an average luminosity and distance of 19,000 $L_{\sun}$ and 6.1 kpc respectively. Additionally, the whole YSO population of the RMS catalogue has averages of 18,000 $L_{\sun}$ and 5.9 kpc. Kolmogorov-Smirnov (K-S) two-sample tests were performed to see whether the single and binary MYSO samples could be deemed to come from the same population. Cumulative distribution plots of luminosity and distance in our sample are shown in \autoref{fig:K-S_dist}. For the luminosity distribution, the K-S statistic was 0.1 and it was judged that there is a 23 per cent chance that the single and binary stars were drawn from the same distribution. The K-S test was also performed with respect to distance to the primary MYSOs, and resulted in a K-S value of 0.14 and a P-value of 0.04, which indicates they are not drawn from the same distribution. Therefore, there appears to be no significant difference in the distribution of primary MYSO luminosity with or without companions, but MYSOs with detected companions are generally closer. This can be explained by the fact that closer objects are generally resolved to a higher degree, and therefore companions with smaller separations are more likely to be found.

Additionally, a K-S test of the $J-K$ colours of the MYSOs resulted in a P-value of 0.26, indicating that the binary and single MYSO primaries share the same distribution. The binaries appear to be slightly less red in general compared to the singles, implying a lesser extinction which may have allowed companions to be detected more easily.

A K-S test for luminosity between the UKIDSS and VVV surveys indicates that there is no significant difference in the luminosity distributions between either survey. When the same test is performed for distance it is apparent that they are not drawn from the same sample; however this may be a result of the different regions of the sky that UKIDSS and VVV probe. The cumulative distribution plots comparing UKIDSS and VVV are shown in \autoref{fig:K-S_ukidssvvv}. As shown in \autoref{fig:galaxy_pos}, UKIDSS targets objects in both the inner and outer galactic spiral arms, with peaks in object frequency at $\sim$1.5 and $\sim$5 kpc. VVV focuses on primarily the inner regions of the galaxy, with a peak at $\sim$3.5 kpc. It is therefore reasonable to assume that this is why the K-S test deems them to have separate distance distributions. This may also explain the different distance distributions between single and binary MYSOs due to the differing background densities (and therefore binary fractions) between the surveys. To conclude, there appear no apparent differences in properties between single and binary YSOs. 

\subsection{Total multiplicity}

Our study significantly improves on the first MYSO multiplicity survey by \citet{POM19}. However, the multiplicity statistics found here are also limited by the observations. The companions found lie at separations ranging from $\sim$10$^3$-10$^5$ au, and companions at smaller separations than this will not be resolved in the UKIDSS, VVV or UKIRT/RMS data due to the spatial resolution. 

To estimate the total multiplicity fraction, unaffected by our observational limitations, Monte Carlo simulations were performed using an artificial binary population applying the same selection effects as the observations. We assume underlying distributions of a lognormal semi-major axis distribution, and a flat eccentricity distribution. We draw the instantaneous orbital properties of the true anomaly, the inclination of the system, and the relative orientation of the system relative to the observer randomly\footnote{Such that the true anomaly is uniformly distributed in time, and the inclination is distributed as sin $i$.}. Using these orbital properties, and the distance distribution of our sample, we can calculate the separation in arcseconds of each simulated companion.  We also draw a magnitude difference between the primary and companion from a truncated normal distribution (truncated at the minimum and maximum observed $\delta$mag values). As the standard deviation tends to higher values, the $\delta$mag distribution becomes flat, allowing us to also include models with a flat uniform $\delta$mag distribution.

We applied the selection effects present in the observed sample to the artificial population, including the gradual decrease in binary detections below $\sim$2 arcseconds and the limiting magnitudes in these regions. We also generated an artificial background density with the same distribution as the observed sample and used this to assign each binary a value of $P_\mathrm{chance}$ (as calculated from \autoref{eq1}). 

\subsubsection{Model results}

\begin{figure}
    \centering
    \begin{subfigure}[t]{0.485\textwidth}
        \centering
        \includegraphics[width=\textwidth]{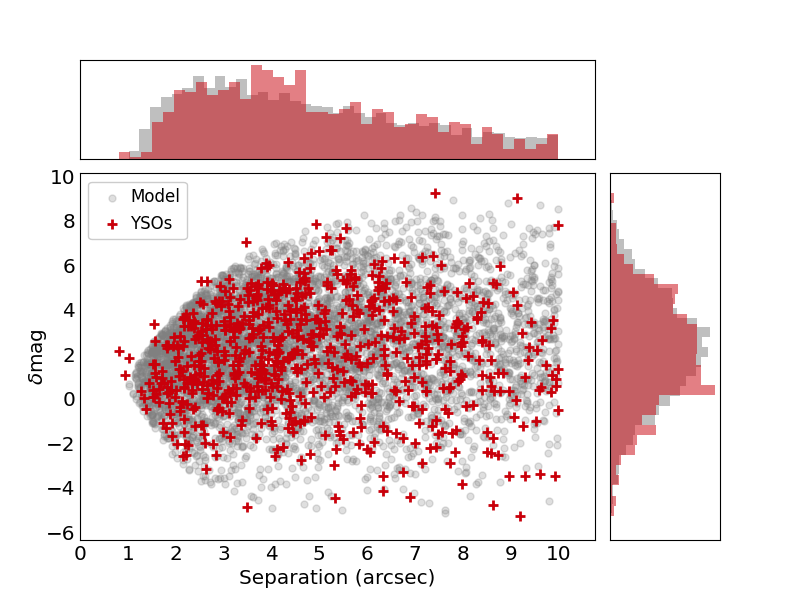}
    \end{subfigure}
    \hfill
    \begin{subfigure}[t]{0.485\textwidth}
        \centering
        \includegraphics[width=\textwidth]{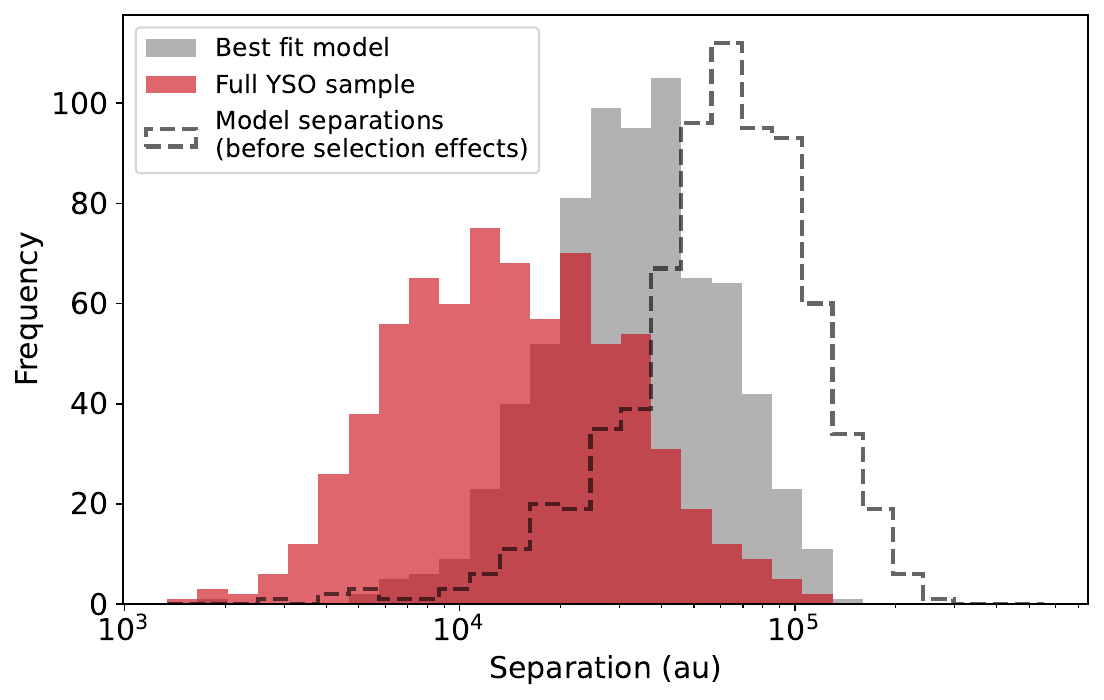}
    \end{subfigure}
    \caption{Top: comparison of the separation vs $\delta$mag distribution from the best fit binary population model (grey) and the combined YSO sample (red). Bottom: the intrinsic log semi-major axis distribution of the binary population from the best fit model (dashed line: intrinsic, grey: simulated distribution) compared to the separation distribution in au from the YSO sample (in red). The three model distributions are normalised such as to fit in the graph. In practice, using the selection criteria, only a few per cent of model systems are observable (see text).}
    \label{fig:selection_effects}
\end{figure}

The results of the models are compared to the observed YSO separation and $\delta$mag distributions in the top of \autoref{fig:selection_effects}.  The main panel shows the observational data from \autoref{fig:dmag} overplotted on the results of $\sim10^4$ simulated systems for the best fitting model. At the top and the side are histograms of the magnitude differences and separations of the simulated (grey) and real (red) data respectively.
The bottom panel shows a histogram of the observed separation distributions as well as those for the intrinsic model separation and the resulting simulated observed model distributions. 

The observed separations peak at $\sim$9000 au, and the best fitting models imply that an extremely wide separation distribution (peaking at $\sim$60,000 au) is required to fit the observations.
This is rather unexpected; we would typically find the peak in the observed separation distribution to be only slightly lower than the peak in the semi-major axis distribution. However, a binary population with a semi-major axis distribution similar to the separation distribution would be much more heavily weighted towards small separations than actually observed. A consequence of the very wide intrinsic separation distribution is that at most 1-3 per cent of all model binaries is ``observed'' due to the observational biases and selection effects. This is a very low number, and much lower than the observed companion fraction of $\sim 30$ per cent.

Hence, it turned out to be extremely difficult to fit the wide companions as the tail of a single binary population - there are just too few that would be observed given the selection effects.  Given that companions are regularly found at 10-100s of au around MYSOs (see Introduction), this suggests that we may be observing triple companions to binaries that are too close to be resolved in the current data, probably combined with a `clustered' component of associated stars ($10^5$ au is a typical distance between stars in a reasonably dense environment such as an association or small cluster).  In some cases where we have multiple companions we are almost certainly seeing a true cluster (several dozens of bound stars), or a loose association/moving group that is still to unbind\footnote{Modelling observations of triples within groups is extremely complex, and possibly too poorly constrained to provide useful results, and is beyond the scope of this paper.}.

To conclude this section, in order to fit the observed data not only do $\sim 100$ per cent of MYSOs need to be in multiples, but a significant fraction of MYSOs (possibly up to 100 per cent) must be in triple systems (see also \citealt{Dodd2023} for a perspective on B stars), and many are still in clusters/associations/moving groups. As it has been previously suggested that up to 100 per cent of massive stars form in binary systems \citep{CHI12}, this work suggests that these objects are frequently found with a higher order of multiplicity than originally thought. Additionally, the high outer Galaxy multiple fraction of 80$^{+6}_{-7}$ per cent mentioned in \autoref{statdiffs} is another indicator that high multiplicity is common in this kind of object.

\section{Conclusions}

We have investigated the binary properties of 681 YSOs (402 of which are MYSOs, i.e. $>$8M$_\odot$) across the RMS catalogue using UKIDSS and VVV point source data, and a sample of 88 YSOs were investigated using $K$-band UKIRT images. Using statistical methods, the probability of companions being real rather than chance projections was used to determine the multiplicity statistics of the sample.
\begin{enumerate}
    \item For the RMS-wide sample using UKIDSS/VVV data, the fractions are MF = $65\pm 4$ per cent and CF = $147\pm 6$ per cent for the UKIDSS sample, and MF = $53\pm 4$ per cent and CF = $84\pm 6$ per cent for the VVV sample. These agree with previous YSO multiplicity studies at similar separation ranges (1000s-10,000s of au).
    \item The multiplicity statistics for the sample of 88 YSOs investigated with the RMS images are MF = $64\pm 8$ per cent and CF = $139\pm 9$ per cent.
    \item A large fraction of companion mass ratios are larger than 0.5, suggesting disagreement with the smaller mass ratios of the binary capture formation scenario.
    \item YSOs in the inner Galaxy have almost identical multiplicity statistics between the UKIDSS/VVV surveys ($\sim$53 per cent). Outer Galaxy YSOs have a multiplicity fraction of 80$^{+6}_{-7}$ per cent. This difference is due to an uneven background density in the UKIDSS survey - and indicates a binary fraction in the probed separation regime close to 100 per cent without the need to consider selection biases.
    \item There appear to be no significant differences in binary and single YSO luminosity and colour, however companions are more likely to be found at closer distances.
    \item The total multiplicity fraction of MYSOs is $\sim$100 per cent, with a large fraction of these (again possibly up to 100 per cent) likely to be at least triple systems, with many associated with clusters/associations/moving groups.
\end{enumerate}
This is one of the first statistical studies, and the largest, specifically dedicated to MYSO multiplicity. Future spectroscopic and interferometric observations will be paramount in learning more about the identified companions, including classifying their spectral types and investigating their environments.

\section*{Acknowledgements}
This work made use of information from the Red MSX Source survey database at \url{http://rms.leeds.ac.uk}, which was constructed with support from the Science and Technology Facilities Council of the UK. It also makes use of the SIMBAD database, operated at CDS, Strasbourg, France. This work uses data obtained as part of the UKIRT Infrared Deep Sky Survey, as well as Vista Variables in the Via Lactea survey data products from observations made with ESO Telescopes at the La Silla or Paranal Observatories under ESO programme ID 179.A-2010. It also uses data from the UKIRT telescope in Hawaii, owned by the University of Hawaii (UH) and operated by the UH Institute for Astronomy. When the data reported here were obtained, UKIRT was operated by the Joint Astronomy Centre on behalf of the Science and Technology Facilities Council of the U.K.

\section*{Data Availability}
The data underlying this article are available in the CDS VizieR database (\url{https://vizier.cds.unistra.fr}).

\bibliography{ref.bib}

\begin{thebibliography}{}
\makeatletter
\relax
\def\mn@urlcharsother{\let\do\@makeother \do\$\do\&\do\#\do\^\do\_\do\%\do\~}
\def\mn@doi{\begingroup\mn@urlcharsother \@ifnextchar [ {\mn@doi@} {\mn@doi@[]}}
\def\mn@doi@[#1]#2{\def\@tempa{#1}\ifx\@tempa\@empty \href {http://dx.doi.org/#2} {doi:#2}\else \href {http://dx.doi.org/#2} {#1}\fi \endgroup}
\def\mn@eprint#1#2{\mn@eprint@#1:#2::\@nil}
\def\mn@eprint@arXiv#1{\href {http://arxiv.org/abs/#1} {{\tt arXiv:#1}}}
\def\mn@eprint@dblp#1{\href {http://dblp.uni-trier.de/rec/bibtex/#1.xml} {dblp:#1}}
\def\mn@eprint@#1:#2:#3:#4\@nil{\def\@tempa {#1}\def\@tempb {#2}\def\@tempc {#3}\ifx \@tempc \@empty \let \@tempc \@tempb \let \@tempb \@tempa \fi \ifx \@tempb \@empty \def\@tempb {arXiv}\fi \@ifundefined {mn@eprint@\@tempb}{\@tempb:\@tempc}{\expandafter \expandafter \csname mn@eprint@\@tempb\endcsname \expandafter{\@tempc}}}

\bibitem[\protect\citeauthoryear{{Astropy Collaboration} et~al.,}{{Astropy Collaboration} et~al.}{2013}]{ASTPY1}
{Astropy Collaboration} et~al., 2013, \mn@doi [\aap] {10.1051/0004-6361/201322068}, \href {http://adsabs.harvard.edu/abs/2013A%26A...558A..33A} {558, A33}

\bibitem[\protect\citeauthoryear{{Badenes} et~al.,}{{Badenes} et~al.}{2018}]{Badenes2018}
{Badenes} C.,  et~al., 2018, \mn@doi [\apj] {10.3847/1538-4357/aaa765}, \href {https://ui.adsabs.harvard.edu/abs/2018ApJ...854..147B} {854, 147}

\bibitem[\protect\citeauthoryear{Baines, Oudmaijer, Porter  \& Pozzo}{Baines et~al.}{2006}]{BAI12}
Baines D.,  Oudmaijer R.~D.,  Porter J.~M.,   Pozzo M.,  2006, \mnras, 367, 737

\bibitem[\protect\citeauthoryear{{Banyard}, {Sana}, {Mahy}, {Bodensteiner}, {Villase{\~n}or}  \& {Evans}}{{Banyard} et~al.}{2022}]{BAN22}
{Banyard} G.,  {Sana} H.,  {Mahy} L.,  {Bodensteiner} J.,  {Villase{\~n}or} J.~I.,   {Evans} C.~J.,  2022, \mn@doi [\aap] {10.1051/0004-6361/202141037}, \href {https://ui.adsabs.harvard.edu/abs/2022A&A...658A..69B} {658, A69}

\bibitem[\protect\citeauthoryear{{Bate}}{{Bate}}{2012}]{BAT12}
{Bate} M.~R.,  2012, \mn@doi [\mnras] {10.1111/j.1365-2966.2011.19955.x}, \href {https://ui.adsabs.harvard.edu/abs/2012MNRAS.419.3115B} {419, 3115}

\bibitem[\protect\citeauthoryear{{Bate}}{{Bate}}{2019}]{Bate2019}
{Bate} M.~R.,  2019, \mn@doi [\mnras] {10.1093/mnras/stz103}, \href {https://ui.adsabs.harvard.edu/abs/2019MNRAS.484.2341B} {484, 2341}

\bibitem[\protect\citeauthoryear{{Beuther} et~al.,}{{Beuther} et~al.}{2019}]{BEU19}
{Beuther} H.,  et~al., 2019, \mn@doi [\aap] {10.1051/0004-6361/201834064}, \href {https://ui.adsabs.harvard.edu/abs/2019A&A...621A.122B} {621, A122}

\bibitem[\protect\citeauthoryear{{Bordier}, {Frost}, {Sana}, {Reggiani}, {M{\'e}rand}, {Rainot}, {Ram{\'\i}rez-Tannus}  \& {de Wit}}{{Bordier} et~al.}{2022}]{BOR22}
{Bordier} E.,  {Frost} A.~J.,  {Sana} H.,  {Reggiani} M.,  {M{\'e}rand} A.,  {Rainot} A.,  {Ram{\'\i}rez-Tannus} M.~C.,   {de Wit} W.~J.,  2022, \mn@doi [\aap] {10.1051/0004-6361/202141849}, \href {https://ui.adsabs.harvard.edu/abs/2022A&A...663A..26B} {663, A26}

\bibitem[\protect\citeauthoryear{{Bressan}, {Marigo}, {Girardi}, {Salasnich}, {Dal Cero}, {Rubele}  \& {Nanni}}{{Bressan} et~al.}{2012}]{Bressan2012}
{Bressan} A.,  {Marigo} P.,  {Girardi} L.,  {Salasnich} B.,  {Dal Cero} C.,  {Rubele} S.,   {Nanni} A.,  2012, \mn@doi [\mnras] {10.1111/j.1365-2966.2012.21948.x}, \href {https://ui.adsabs.harvard.edu/abs/2012MNRAS.427..127B} {427, 127}

\bibitem[\protect\citeauthoryear{{Capitanio}, {Lallement}, {Vergely}, {Elyajouri}  \& {Monreal-Ibero}}{{Capitanio} et~al.}{2017}]{STILISM17}
{Capitanio} L.,  {Lallement} R.,  {Vergely} J.~L.,  {Elyajouri} M.,   {Monreal-Ibero} A.,  2017, \mn@doi [\aap] {10.1051/0004-6361/201730831}, \href {https://ui.adsabs.harvard.edu/abs/2017A&A...606A..65C} {606, A65}

\bibitem[\protect\citeauthoryear{{Chini}, {Hoffmeister}, {Nasseri}, {Stahl}  \& {Zinnecker}}{{Chini} et~al.}{2012}]{CHI12}
{Chini} R.,  {Hoffmeister} V.~H.,  {Nasseri} A.,  {Stahl} O.,   {Zinnecker} H.,  2012, \mn@doi [\mnras] {10.1111/j.1365-2966.2012.21317.x}, \href {https://ui.adsabs.harvard.edu/abs/2012MNRAS.424.1925C} {424, 1925}

\bibitem[\protect\citeauthoryear{{Clarke}, {Lumsden}, {Oudmaijer}, {Busfield}, {Hoare}, {Moore}, {Sheret}  \& {Urquhart}}{{Clarke} et~al.}{2006}]{CLA06}
{Clarke} A.~J.,  {Lumsden} S.~L.,  {Oudmaijer} R.~D.,  {Busfield} A.~L.,  {Hoare} M.~G.,  {Moore} T.~J.~T.,  {Sheret} T.~L.,   {Urquhart} J.~S.,  2006, \mn@doi [\aap] {10.1051/0004-6361:20064839}, \href {https://ui.adsabs.harvard.edu/abs/2006A&A...457..183C} {457, 183}

\bibitem[\protect\citeauthoryear{Cooper}{Cooper}{2013}]{COOTHESIS}
Cooper H. D.~B.,  2013, Thesis, University of Leeds, \url {https://ui.adsabs.harvard.edu/abs/2013PhDT.......441C}

\bibitem[\protect\citeauthoryear{{Cooper} et~al.,}{{Cooper} et~al.}{2013}]{COO13}
{Cooper} H.~D.~B.,  et~al., 2013, \mn@doi [\mnras] {10.1093/mnras/sts681}, \href {https://ui.adsabs.harvard.edu/abs/2013MNRAS.430.1125C} {430, 1125}

\bibitem[\protect\citeauthoryear{{Correia}, {Zinnecker}, {Ratzka}  \& {Sterzik}}{{Correia} et~al.}{2006}]{COR06}
{Correia} S.,  {Zinnecker} H.,  {Ratzka} T.,   {Sterzik} M.~F.,  2006, \mn@doi [\aap] {10.1051/0004-6361:20065545}, \href {https://ui.adsabs.harvard.edu/abs/2006A&A...459..909C} {459, 909}

\bibitem[\protect\citeauthoryear{{Cyganowski}, {Ilee}, {Brogan}, {Hunter}, {Zhang}, {Harries}  \& {Haworth}}{{Cyganowski} et~al.}{2022}]{Cyganowski2022}
{Cyganowski} C.~J.,  {Ilee} J.~D.,  {Brogan} C.~L.,  {Hunter} T.~R.,  {Zhang} S.,  {Harries} T.~J.,   {Haworth} T.~J.,  2022, \mn@doi [\apjl] {10.3847/2041-8213/ac69ca}, \href {https://ui.adsabs.harvard.edu/abs/2022ApJ...931L..31C} {931, L31}

\bibitem[\protect\citeauthoryear{{Davies}, {Hoare}, {Lumsden}, {Hosokawa}, {Oudmaijer}, {Urquhart}, {Mottram}  \& {Stead}}{{Davies} et~al.}{2011}]{DAV11}
{Davies} B.,  {Hoare} M.~G.,  {Lumsden} S.~L.,  {Hosokawa} T.,  {Oudmaijer} R.~D.,  {Urquhart} J.~S.,  {Mottram} J.~C.,   {Stead} J.,  2011, \mn@doi [\mnras] {10.1111/j.1365-2966.2011.19095.x}, \href {https://ui.adsabs.harvard.edu/abs/2011MNRAS.416..972D} {416, 972}

\bibitem[\protect\citeauthoryear{{Dodd}, {Oudmaijer}, {Radley}, {Vioque}  \& {Frost}}{{Dodd} et~al.}{2023}]{Dodd2023}
{Dodd} J.~M.,  {Oudmaijer} R.~D.,  {Radley} I.~C.,  {Vioque} M.,   {Frost} A.~J.,  2023, \mn@doi [\mnras] {10.1093/mnras/stad3105}, \href {https://ui.adsabs.harvard.edu/abs/2023MNRAS.tmp.2984D} {}

\bibitem[\protect\citeauthoryear{{Duch{\^e}ne} \& {Kraus}}{{Duch{\^e}ne} \& {Kraus}}{2013}]{DUC13}
{Duch{\^e}ne} G.,  {Kraus} A.,  2013, \mn@doi [\araa] {10.1146/annurev-astro-081710-102602}, \href {https://ui.adsabs.harvard.edu/abs/2013ARA&A..51..269D} {51, 269}

\bibitem[\protect\citeauthoryear{{Frost}, {Oudmaijer}, {de Wit}  \& {Lumsden}}{{Frost} et~al.}{2019}]{FRO19}
{Frost} A.~J.,  {Oudmaijer} R.~D.,  {de Wit} W.~J.,   {Lumsden} S.~L.,  2019, \mn@doi [\aap] {10.1051/0004-6361/201834583}, \href {https://ui.adsabs.harvard.edu/abs/2019A&A...625A..44F} {625, A44}

\bibitem[\protect\citeauthoryear{{Frost}, {Oudmaijer}, {de Wit}  \& {Lumsden}}{{Frost} et~al.}{2021}]{FRO21}
{Frost} A.~J.,  {Oudmaijer} R.~D.,  {de Wit} W.~J.,   {Lumsden} S.~L.,  2021, \mn@doi [\aap] {10.1051/0004-6361/202039748}, \href {https://ui.adsabs.harvard.edu/abs/2021A&A...648A..62F} {648, A62}

\bibitem[\protect\citeauthoryear{{Green}, {Schlafly}, {Zucker}, {Speagle}  \& {Finkbeiner}}{{Green} et~al.}{2019}]{BAYESTAR19}
{Green} G.~M.,  {Schlafly} E.,  {Zucker} C.,  {Speagle} J.~S.,   {Finkbeiner} D.,  2019, \mn@doi [\apj] {10.3847/1538-4357/ab5362}, \href {https://ui.adsabs.harvard.edu/abs/2019ApJ...887...93G} {887, 93}

\bibitem[\protect\citeauthoryear{{Griffiths}, {Goodwin}  \& {Caballero-Nieves}}{{Griffiths} et~al.}{2018}]{GRI18}
{Griffiths} D.~W.,  {Goodwin} S.~P.,   {Caballero-Nieves} S.~M.,  2018, \mn@doi [\mnras] {10.1093/mnras/sty412}, \href {https://ui.adsabs.harvard.edu/abs/2018MNRAS.476.2493G} {476, 2493}

\bibitem[\protect\citeauthoryear{{Halbwachs}}{{Halbwachs}}{1988}]{Halbwachs1988}
{Halbwachs} J.~L.,  1988, \mn@doi [\apss] {10.1007/BF00656216}, \href {https://ui.adsabs.harvard.edu/abs/1988Ap&SS.142..237H} {142, 237}

\bibitem[\protect\citeauthoryear{{Harada}, {Hirano}, {Machida}  \& {Hosokawa}}{{Harada} et~al.}{2021}]{Harada2021}
{Harada} N.,  {Hirano} S.,  {Machida} M.~N.,   {Hosokawa} T.,  2021, \mn@doi [\mnras] {10.1093/mnras/stab2780}, \href {https://ui.adsabs.harvard.edu/abs/2021MNRAS.508.3730H} {508, 3730}

\bibitem[\protect\citeauthoryear{{Kennicutt}}{{Kennicutt}}{2005}]{KEN05}
{Kennicutt} R.~C.,  2005, in {Cesaroni} R.,  {Felli} M.,  {Churchwell} E.,   {Walmsley} M.,  eds,  IAU Symposium Vol. 227, Massive Star Birth: A Crossroads of Astrophysics. pp 3--11, \mn@doi{10.1017/S1743921305004308}

\bibitem[\protect\citeauthoryear{{King}, {Parker}, {Patience}  \& {Goodwin}}{{King} et~al.}{2012}]{KIN12}
{King} R.~R.,  {Parker} R.~J.,  {Patience} J.,   {Goodwin} S.~P.,  2012, \mn@doi [\mnras] {10.1111/j.1365-2966.2012.20437.x}, \href {https://ui.adsabs.harvard.edu/abs/2012MNRAS.421.2025K} {421, 2025}

\bibitem[\protect\citeauthoryear{{Koumpia} et~al.,}{{Koumpia} et~al.}{2019}]{KOU19}
{Koumpia} E.,  et~al., 2019, \mn@doi [\aap] {10.1051/0004-6361/201834624}, \href {https://ui.adsabs.harvard.edu/abs/2019A&A...623L...5K} {623, L5}

\bibitem[\protect\citeauthoryear{{Koumpia} et~al.,}{{Koumpia} et~al.}{2021}]{KOU21}
{Koumpia} E.,  et~al., 2021, \mn@doi [\aap] {10.1051/0004-6361/202141373}, \href {https://ui.adsabs.harvard.edu/abs/2021A&A...654A.109K} {654, A109}

\bibitem[\protect\citeauthoryear{{Kratter}}{{Kratter}}{2011}]{KRA11}
{Kratter} K.~M.,  2011, in {Schmidtobreick} L.,  {Schreiber} M.~R.,   {Tappert} C.,  eds,  Astronomical Society of the Pacific Conference Series Vol. 447, Evolution of Compact Binaries. p.~47 (\mn@eprint {arXiv} {1109.3740})

\bibitem[\protect\citeauthoryear{{Kratter}, {Matzner}  \& {Krumholz}}{{Kratter} et~al.}{2008}]{KRA08}
{Kratter} K.~M.,  {Matzner} C.~D.,   {Krumholz} M.~R.,  2008, \mn@doi [\apj] {10.1086/587543}, \href {https://ui.adsabs.harvard.edu/abs/2008ApJ...681..375K} {681, 375}

\bibitem[\protect\citeauthoryear{Kraus et~al.,}{Kraus et~al.}{2017}]{KRA17}
Kraus S.,  et~al., 2017, \apj, 835

\bibitem[\protect\citeauthoryear{{Krumholz}, {Klein}, {McKee}, {Offner}  \& {Cunningham}}{{Krumholz} et~al.}{2009}]{KRU09}
{Krumholz} M.~R.,  {Klein} R.~I.,  {McKee} C.~F.,  {Offner} S. S.~R.,   {Cunningham} A.~J.,  2009, \mn@doi [\sci] {10.1126/science.1165857}, \href {https://ui.adsabs.harvard.edu/abs/2009Sci...323..754K} {323, 754}

\bibitem[\protect\citeauthoryear{{Lucas} et~al.,}{{Lucas} et~al.}{2008}]{UKIDSS}
{Lucas} P.~W.,  et~al., 2008, \mn@doi [\mnras] {10.1111/j.1365-2966.2008.13924.x}, \href {https://ui.adsabs.harvard.edu/abs/2008MNRAS.391..136L} {391, 136}

\bibitem[\protect\citeauthoryear{{Lumsden}, {Hoare}, {Urquhart}, {Oudmaijer}, {Davies}, {Mottram}, {Cooper}  \& {Moore}}{{Lumsden} et~al.}{2013}]{LUM13}
{Lumsden} S.~L.,  {Hoare} M.~G.,  {Urquhart} J.~S.,  {Oudmaijer} R.~D.,  {Davies} B.,  {Mottram} J.~C.,  {Cooper} H.~D.~B.,   {Moore} T.~J.~T.,  2013, \mn@doi [\apjs] {10.1088/0067-0049/208/1/11}, \href {https://ui.adsabs.harvard.edu/abs/2013ApJS..208...11L} {208, 11}

\bibitem[\protect\citeauthoryear{{Lund} \& {Bonnell}}{{Lund} \& {Bonnell}}{2018}]{LUN18}
{Lund} K.,  {Bonnell} I.~A.,  2018, \mn@doi [\mnras] {10.1093/mnras/sty1584}, \href {https://ui.adsabs.harvard.edu/abs/2018MNRAS.479.2235L} {479, 2235}

\bibitem[\protect\citeauthoryear{{Marks} \& {Kroupa}}{{Marks} \& {Kroupa}}{2012}]{MAR12}
{Marks} M.,  {Kroupa} P.,  2012, \mn@doi [\aap] {10.1051/0004-6361/201118231}, \href {https://ui.adsabs.harvard.edu/abs/2012A&A...543A...8M} {543, A8}

\bibitem[\protect\citeauthoryear{{Mathieu}}{{Mathieu}}{1994}]{MAT94}
{Mathieu} R.~D.,  1994, \mn@doi [\araa] {10.1146/annurev.aa.32.090194.002341}, \href {https://ui.adsabs.harvard.edu/abs/1994ARA&A..32..465M} {32, 465}

\bibitem[\protect\citeauthoryear{{M{\'e}ndez-Delgado}, {Amayo}, {Arellano-C{\'o}rdova}, {Esteban}, {Garc{\'\i}a-Rojas}, {Carigi}  \& {Delgado-Inglada}}{{M{\'e}ndez-Delgado} et~al.}{2022}]{Mendez2022}
{M{\'e}ndez-Delgado} J.~E.,  {Amayo} A.,  {Arellano-C{\'o}rdova} K.~Z.,  {Esteban} C.,  {Garc{\'\i}a-Rojas} J.,  {Carigi} L.,   {Delgado-Inglada} G.,  2022, \mn@doi [\mnras] {10.1093/mnras/stab3782}, \href {https://ui.adsabs.harvard.edu/abs/2022MNRAS.510.4436M} {510, 4436}

\bibitem[\protect\citeauthoryear{{Meyer}, {Kuiper}, {Kley}, {Johnston}  \& {Vorobyov}}{{Meyer} et~al.}{2018}]{MEY18}
{Meyer} D.~M.~A.,  {Kuiper} R.,  {Kley} W.,  {Johnston} K.~G.,   {Vorobyov} E.,  2018, \mn@doi [\mnras] {10.1093/mnras/stx2551}, \href {https://ui.adsabs.harvard.edu/abs/2018MNRAS.473.3615M} {473, 3615}

\bibitem[\protect\citeauthoryear{{Meyer}, {Kreplin}, {Kraus}, {Vorobyov}, {Haemmerle}  \& {Eisl{\"o}ffel}}{{Meyer} et~al.}{2019}]{MEY19}
{Meyer} D.~M.~A.,  {Kreplin} A.,  {Kraus} S.,  {Vorobyov} E.~I.,  {Haemmerle} L.,   {Eisl{\"o}ffel} J.,  2019, \mn@doi [\mnras] {10.1093/mnras/stz1585}, \href {https://ui.adsabs.harvard.edu/abs/2019MNRAS.487.4473M} {487, 4473}

\bibitem[\protect\citeauthoryear{{Moe} \& {Di Stefano}}{{Moe} \& {Di Stefano}}{2017}]{MOE17}
{Moe} M.,  {Di Stefano} R.,  2017, \mn@doi [\apjs] {10.3847/1538-4365/aa6fb6}, \href {https://ui.adsabs.harvard.edu/abs/2017ApJS..230...15M} {230, 15}

\bibitem[\protect\citeauthoryear{{Moe}, {Kratter}  \& {Badenes}}{{Moe} et~al.}{2019}]{Moe2019}
{Moe} M.,  {Kratter} K.~M.,   {Badenes} C.,  2019, \mn@doi [\apj] {10.3847/1538-4357/ab0d88}, \href {https://ui.adsabs.harvard.edu/abs/2019ApJ...875...61M} {875, 61}

\bibitem[\protect\citeauthoryear{{Moeckel} \& {Bate}}{{Moeckel} \& {Bate}}{2010}]{MOE10}
{Moeckel} N.,  {Bate} M.~R.,  2010, \mn@doi [\mnras] {10.1111/j.1365-2966.2010.16347.x}, \href {https://ui.adsabs.harvard.edu/abs/2010MNRAS.404..721M} {404, 721}

\bibitem[\protect\citeauthoryear{{Mottram} et~al.,}{{Mottram} et~al.}{2011}]{MOT11}
{Mottram} J.~C.,  et~al., 2011, \mn@doi [\aap] {10.1051/0004-6361/201014479}, \href {https://ui.adsabs.harvard.edu/abs/2011A&A...525A.149M} {525, A149}

\bibitem[\protect\citeauthoryear{{Myers}, {McKee}, {Cunningham}, {Klein}  \& {Krumholz}}{{Myers} et~al.}{2013}]{MYE13}
{Myers} A.~T.,  {McKee} C.~F.,  {Cunningham} A.~J.,  {Klein} R.~I.,   {Krumholz} M.~R.,  2013, \mn@doi [\apj] {10.1088/0004-637X/766/2/97}, \href {https://ui.adsabs.harvard.edu/abs/2013ApJ...766...97M} {766, 97}

\bibitem[\protect\citeauthoryear{{Offner}, {Moe}, {Kratter}, {Sadavoy}, {Jensen}  \& {Tobin}}{{Offner} et~al.}{2023}]{OFF23}
{Offner} S.~S.~R.,  {Moe} M.,  {Kratter} K.~M.,  {Sadavoy} S.~I.,  {Jensen} E.~L.~N.,   {Tobin} J.~J.,  2023, in {Inutsuka} S.,  {Aikawa} Y.,  {Muto} T.,  {Tomida} K.,   {Tamura} M.,  eds,  Astronomical Society of the Pacific Conference Series Vol. 534, Protostars and Planets VII. p.~275 (\mn@eprint {arXiv} {2203.10066}), \mn@doi{10.48550/arXiv.2203.10066}

\bibitem[\protect\citeauthoryear{{Oudmaijer} \& {Parr}}{{Oudmaijer} \& {Parr}}{2010}]{OUD10}
{Oudmaijer} R.~D.,  {Parr} A.~M.,  2010, \mn@doi [\mnras] {10.1111/j.1365-2966.2010.16609.x}, \href {https://ui.adsabs.harvard.edu/abs/2010MNRAS.405.2439O} {405, 2439}

\bibitem[\protect\citeauthoryear{{Pomohaci}, {Oudmaijer}  \& {Goodwin}}{{Pomohaci} et~al.}{2019}]{POM19}
{Pomohaci} R.,  {Oudmaijer} R.~D.,   {Goodwin} S.~P.,  2019, \mn@doi [\mnras] {10.1093/mnras/stz014}, \href {https://ui.adsabs.harvard.edu/abs/2019MNRAS.484..226P} {484, 226}

\bibitem[\protect\citeauthoryear{{Price-Whelan} et~al.,}{{Price-Whelan} et~al.}{2018}]{ASTPY2}
{Price-Whelan} A.~M.,  et~al., 2018, \mn@doi [\aj] {10.3847/1538-3881/aabc4f}, \href {https://ui.adsabs.harvard.edu/#abs/2018AJ....156..123T} {156, 123}

\bibitem[\protect\citeauthoryear{{Ram{\'\i}rez-Tannus} et~al.,}{{Ram{\'\i}rez-Tannus} et~al.}{2021}]{RAM21}
{Ram{\'\i}rez-Tannus} M.~C.,  et~al., 2021, \mn@doi [\aap] {10.1051/0004-6361/202039673}, \href {https://ui.adsabs.harvard.edu/abs/2021A&A...645L..10R} {645, L10}

\bibitem[\protect\citeauthoryear{{Rosen}, {Li}, {Zhang}  \& {Burkhart}}{{Rosen} et~al.}{2019}]{ROS19}
{Rosen} A.~L.,  {Li} P.~S.,  {Zhang} Q.,   {Burkhart} B.,  2019, \mn@doi [\apj] {10.3847/1538-4357/ab54c6}, \href {https://ui.adsabs.harvard.edu/abs/2019ApJ...887..108R} {887, 108}

\bibitem[\protect\citeauthoryear{{Saito} et~al.,}{{Saito} et~al.}{2012}]{VVV}
{Saito} R.~K.,  et~al., 2012, \mn@doi [\aap] {10.1051/0004-6361/201118407}, \href {https://ui.adsabs.harvard.edu/abs/2012A&A...537A.107S} {537, A107}

\bibitem[\protect\citeauthoryear{{Salpeter}}{{Salpeter}}{1955}]{SAL55}
{Salpeter} E.~E.,  1955, \mn@doi [\apj] {10.1086/145971}, \href {https://ui.adsabs.harvard.edu/abs/1955ApJ...121..161S} {121, 161}

\bibitem[\protect\citeauthoryear{{Sana} et~al.,}{{Sana} et~al.}{2012}]{SAN12}
{Sana} H.,  et~al., 2012, \mn@doi [Science] {10.1126/science.1223344}, \href {https://ui.adsabs.harvard.edu/abs/2012Sci...337..444S} {337, 444}

\bibitem[\protect\citeauthoryear{{Sana} et~al.,}{{Sana} et~al.}{2014}]{SAN14}
{Sana} H.,  et~al., 2014, \mn@doi [\apjs] {10.1088/0067-0049/215/1/15}, \href {https://ui.adsabs.harvard.edu/abs/2014ApJS..215...15S} {215, 15}

\bibitem[\protect\citeauthoryear{{Stetson}}{{Stetson}}{1987}]{STE87}
{Stetson} P.~B.,  1987, \mn@doi [\pasp] {10.1086/131977}, \href {https://ui.adsabs.harvard.edu/abs/1987PASP...99..191S} {99, 191}

\bibitem[\protect\citeauthoryear{{Van Albada}}{{Van Albada}}{1968}]{Albada1968}
{Van Albada} T.~S.,  1968, \bain, \href {https://ui.adsabs.harvard.edu/abs/1968BAN....20...47V} {20, 47}

\bibitem[\protect\citeauthoryear{{Wheelwright}, {Oudmaijer}  \& {Goodwin}}{{Wheelwright} et~al.}{2010}]{WHE10}
{Wheelwright} H.~E.,  {Oudmaijer} R.~D.,   {Goodwin} S.~P.,  2010, \mn@doi [\mnras] {10.1111/j.1365-2966.2009.15708.x}, \href {https://ui.adsabs.harvard.edu/abs/2010MNRAS.401.1199W} {401, 1199}

\bibitem[\protect\citeauthoryear{{Zhang} et~al.,}{{Zhang} et~al.}{2019}]{ZHA19}
{Zhang} Y.,  et~al., 2019, \mn@doi [Nature Astronomy] {10.1038/s41550-019-0718-y}, \href {https://ui.adsabs.harvard.edu/abs/2019NatAs...3..517Z} {3, 517}

\makeatother
\end{thebibliography}
\appendix
\section{Data Tables}
\begin{table*}
\caption{Table of all primary YSOs studied in this paper. All YSOs were retrieved from the RMS catalogue. The distance and bolometric luminosity (L$_{\mathrm{bol}}$) were retrieved from the RMS database and the masses were computed using the mass-luminosity relation of \citet{DAV11}. The infrared sky survey used to study the YSO is shown in the 'Survey' column. The $JHK$ magnitudes of the primary are taken from either the infrared survey, or 2MASS if the object is likely to be saturated in the infrared survey. Some columns have been omitted here; the full table can be found online.}
\centering
\label{appA}
\begin{tabular}{lllcrrcccc}
\hline
    RMS ID &       RA &      Dec &  Distance &  L$_{\mathrm{bol}}$ &  Mass & Survey &  $J$ &  $H$ &  $K$ \\
    & (deg) & (deg) & (kpc) & (L$_\odot$) & (M$_\odot$) & & (mag) & (mag) & (mag)\\
\hline
    G010.3208-00.1570B & 272.2562 & -20.0856 &   3.5 & 41620 &  20.4 & UKIDSS &      16.7 &      17.2 &      13.6 \\
    G010.3844+02.2128 & 270.0944 & -18.8694 &   1.1 &  1180 &   6.3 & UKIDSS &      16.5 &      14.0 &      10.5 \\
    G010.5067+02.2285 & 270.1439 &  -18.755 &   2.9 &  1660 &   7.0 & UKIDSS &           &      16.7 &      14.1 \\
    G010.8856+00.1221 & 272.2833 & -19.4567 &   2.7 &  3560 &   8.8 & UKIDSS &      18.6 &      13.2 &       9.6 \\
    G011.4201-01.6815 & 274.2362 & -19.8522 &   1.5 &  7040 &  11.0 & UKIDSS &      18.9 &      14.5 &      11.7 \\
    ... & ... & ... & ... & ... & ... & ... & ... & ... & ...\\
\hline
\end{tabular}
\end{table*}

\begin{table*}
\caption{Table of all companions detected using infrared imaging surveys. Companions detected around primaries down to G229.5711+00.1525 were detected in UKIDSS; objects afterwards were detected in VVV (this is highlighted in the 'Survey' column. The $J$,$H$ and $K$ magnitudes are from the corresponding IR survey unless they are brighter than that survey's saturation limit; in these cases 2MASS magnitudes were used instead. q$_{\mathrm{fg,X}}$ represents a mass ratio derived using foreground extinction, and q$_{\mathrm{tot,X}}$ represents a mass ratio derived using total extinction, labelled with the waveband $X$. Some columns have been omitted here; the full table can be found online.}
\centering
\label{appB}
\begin{tabular}{llllcrcccc}
\hline
    Survey ID & RA & Dec & Centre RMS ID & Separation & P$_{\mathrm{chance}}$ 
    & q$_{\mathrm{fg,K}}$ & q$_{\mathrm{fg,J}}$ & q$_{\mathrm{tot,K}}$ & q$_{\mathrm{tot,J}}$\\
    & (deg) & (deg) & & (arcsec) & (\%) & & & &\\
\hline
438306049182 & 270.1444 & -18.7559 & G010.5067+02.2285 &         3.7 &       5.0 &   0.9 &     0.2 &      3.0 &        2.3 \\
438306049183 & 270.1453 & -18.7547 & G010.5067+02.2285 &         4.7 &      13.9 &   0.6 &     0.4 &      0.9 &        0.8 \\
438466784310 & 272.2851 & -19.4582 & G010.8856+00.1221 &         8.8 &      13.5 &   1.0 &     0.2 &      5.2 &        5.5 \\
438466784296 & 272.2841 & -19.4546 & G010.8856+00.1221 &         7.7 &      11.8 &   1.0 &     0.4 &      2.3 &        2.3 \\
438466784158 & 272.2837 & -19.4573 & G010.8856+00.1221 &         3.2 &      13.0 &   0.5 &     0.2 &      1.0 &        1.1 \\
... & ... & ... & ... & ... & ... & ... & ... & ... & ...\\
\hline
\end{tabular}
\end{table*}

\begin{table*}
\caption{Table of all companions detected using source detection in RMS images. The $K-$band magnitudes in this sample were obtained through the source detection program and were then flux-calibrated. The full table can be found online.}
\centering
\label{appC}
\begin{tabular}{lllcrr}
\hline
    RA & Dec & Centre RMS ID & Separation & $\delta K$ & P$_{\mathrm{chance}}$ \\
    (deg) & (deg) & & (arcsec) & (mag) & (\%)\\
\hline
          273.5885 & -12.7422 & G017.3765+02.2512 &         2.8 &  16.1 &       7.0 \\
      278.3774 &  -5.0172 & G026.4207+01.6858 &         0.8 &  15.7 &       0.2 \\
      278.3771 &  -5.0172 & G026.4207+01.6858 &         0.6 &  15.8 &       0.2 \\
      278.3774 &  -5.0176 & G026.4207+01.6858 &         1.0 &  16.7 &       0.9 \\
      284.7796 &   7.0496 & G040.0809+01.5117 &         3.2 &  16.4 &       2.5 \\
      ... & ... & ... & ... & ... & ...\\
\hline
\end{tabular}
\end{table*}
\end{document}